%% file: AcceptedVersion.tex
\documentclass{ieeeaccess}
\usepackage{cite}
\usepackage{amsmath,amssymb,amsfonts}
\usepackage{algorithmic}
\usepackage{graphicx}
\usepackage{textcomp}
\usepackage[nolist]{acronym}
\usepackage{subfigure}
\usepackage{color}
\usepackage{multirow}
\usepackage[normalem]{ulem}
\usepackage{soul}
\usepackage{hyperref}
\usepackage{caption}

\input{definitions.tex}

\input{acronyms.tex}

\def\BibTeX{{\rm B\kern-.05em{\sc i\kern-.025em b}\kern-.08em
    T\kern-.1667em\lower.7ex\hbox{E}\kern-.125emX}}
\begin{document}
\history{Received October 30, 2018, accepted November 19, 2018. Date of publication xxxx 00, 0000, date of current version xxxx 00, 0000.}
\doi{10.1109/ACCESS.2018.2883401}

\title{Study of the Impact of PHY and MAC \\Parameters in 3GPP C-V2V Mode 4}
\author{\uppercase{Alessandro Bazzi}, \IEEEmembership{Senior Member, IEEE},\\
\uppercase{Giammarco Cecchini}, \IEEEmembership{Student Member, IEEE},\\
\uppercase{Alberto Zanella}, \IEEEmembership{Senior Member, IEEE},\\
\uppercase{Barbara M. Masini}, \IEEEmembership{Member, IEEE},
}
\address[1]{CNR-IEIIT, Bologna, Italy (e-mail: \{alessandro.bazzi, giammarco.cecchini,  alberto.zanella, barbara.masini\}@ieiit.cnr.it)}

\markboth
{A. Bazzi \headeretal: Study of the Impact of PHY and MAC Parameters in 3GPP C-V2V Mode 4}
{A. Bazzi \headeretal: Study of the Impact of PHY and MAC Parameters in 3GPP C-V2V Mode 4}

\corresp{Corresponding author: Alessandro Bazzi (e-mail: alessandro.bazzi@ieiit.cnr.it).\\ \vskip 0.1cm
This is the accepted version of the paper. The published manuscript
may undergo copyediting, typesetting, and review before it is published
in its final form. A definitive version will be subsequently published.\\\textcopyright 2018 IEEE. Translations and content mining are permitted for academic research only.
Personal use is also permitted, but republication/redistribution requires IEEE permission.
See http://www.ieee.org/publications\_standards/publications/rights/index.html for more information.}

\begin{abstract}
In the latest years, 3GPP has added short-range \ac{C-V2X} to the features of LTE and 5G to allow vehicles, roadside devices, and vulnerable users  to directly exchange information using the same chipset as for classical long-range connections. C-V2X is based on the use of advanced physical layer techniques and orthogonal resources, and one of the main aspects affecting its performance is the way resources are allocated. Allocations  can be either managed by the network or in a distributed way, directly by the nodes. The latter case, called Mode 4, is required in those situations where the network cannot be involved in the scheduling process, for example due to a lack of coverage, but could also be adopted in order to reduce the processing burden of eNodeB. An algorithm, defined in the standards, makes nodes sense the medium and identify the best time-frequency combination to allocate their messages. Focusing on C-V2X Mode 4, in this work we analyse the parameters of the algorithm designed by 3GPP and their impact on the system performance. Through simulations in different large-scale scenarios, we show that modifying some parameters have negligible effect, that the proper choice of others can indeed improve the quality of service, and that a group of parameters allows to trade-off reliability with update delay. The provided results can also be exploited to guide future work. 
\end{abstract}

\begin{keywords}
C-V2X, Intelligent vehicles, Vehicular and wireless technologies, Wireless networks
\end{keywords}

\titlepgskip=-15pt

\maketitle

\acresetall

\section{Introduction}\label{Section:intro}

Everyone is agreed that \acp{CAV} are coming on our roads in the next few years, completely changing the way mobility is conceived today. As often remarked, full automation is not enough and wireless technologies will play a key role. 

In this scenario, as an alternative to the well known and widely tested IEEE 802.11p (and related standards), 3GPP has added new dedicated features to the cellular  ecosystem to enable short range communications in the so-called \ac{C-V2X}. More specifically, by the end of 2016, advanced features have been added in Release 14 to enable direct \ac{D2D} communications for the specific scenario of vehicular networks \cite{BazMasZanThi:J17}. Such technology will enable short-range \ac{V2V}, \ac{V2I}, and \ac{V2P} communications, integrated with classical long-range coverage.

Distinctive characteristics of short-range C-V2X with respect to IEEE 802.11p are its advanced \ac{PHY} layer and the use of orthogonal resources at the \ac{MAC} layer. In principle, different users can transmit in fully separate time-frequency slots, thus avoiding reciprocal interference. However, this beneficial effect strictly depends on the ability to identify those resources that are not occupied. In addition, the available spectrum is scarce and space diversity becomes a third optimization dimension in order to minimize the reciprocal interference.

In short-range C-V2X, two different approaches are defined for resource allocation, one under the control of the network, called Mode~3, and the other where decisions are fully distributed among nodes, denoted as Mode~4. Although Mode~3 is expected to outperform Mode~4 given the more information available at the scheduler \cite{MinWinZhaBlaEtc:C17,CecBazMasZan:C17_4}, still problems could arise at the cellular boundaries (especially with different operators) and the latter remains the only option when coverage is intermittent or not available.

Given its crucial importance and in order to make products from  different vendors interoperable, a Mode 4 algorithm is defined by 3GPP in \cite{3GPP_TS_36_213,3GPP_TS_36_321}. Outside 3GPP, still few works have investigated its performance and the impact of the various parameters. A study is presented for example in  \cite{MolGoz:C17}, where the authors compare Mode~4 to a random selection of resources. Results, in terms of delivery rate, are provided in a Manhattan grid scenario. The same authors extend the investigation to a highway environment in \cite{MolGoz:J17},   while in \cite{GonSepMolGoz:AX18} an analytical model is proposed in simplified scenarios.  Furthermore, in \cite{NguShaSudKapEtc:C17} Nguyen et al. compare Mode~4 with a controlled allocation scheme and with IEEE 802.11p. All these works assume fixed and arbitrary settings of the parameters. Some very recent works have focused on the impact of the parameters, limiting their studies to specific aspects and showing that the settings can significantly affect the performance \cite{NabMarKauDie:AX18,TogSaiMugMahetal:AX18}.

Although several papers have recently concentrated on Mode~4, they have all posed the attention on specific aspects and one or few parameters. To cope with this limitation and to provide an in-depth discussion of Mode~4, here we focus on the main parameters defined at the \ac{PHY} and \ac{MAC} layers, considering both those that can be adjusted by specification and those that are instead mandate to a given value. The study will focus on the \ac{SPS} related to the cooperative awareness service, which is the periodic broadcast of updated information by all vehicles about their status and movements and is at the basis of most of the applications foreseen for connected vehicles \cite{LyaVinJonBel:J18,BalUhlCalCan:J18}. The impact of each parameter and its optimal setting is derived through large-scale simulations in three realistic scenarios with hundred of nodes, using LTEV2Vsim \cite{CecBazMasZan:C17}. 

In summary, the contribution of the paper is as follows.
\begin{itemize}
	\item The performance of C-V2X Mode~4 is shown in realistic urban and highway scenarios, varying all main parameters at PHY and MAC layers;
	\item We evaluate the impact of each parameter, starting from, but not constrained to, the values indicated by 3GPP, and identify the optimal settings;
	\item Based on the results, guidelines for future improvements of the C-V2X Mode~4 algorithm are provided.
\end{itemize}

\begin{figure*}[t]
	\centering
	\includegraphics[trim={50 340 200 20},clip,width=0.95\textwidth]{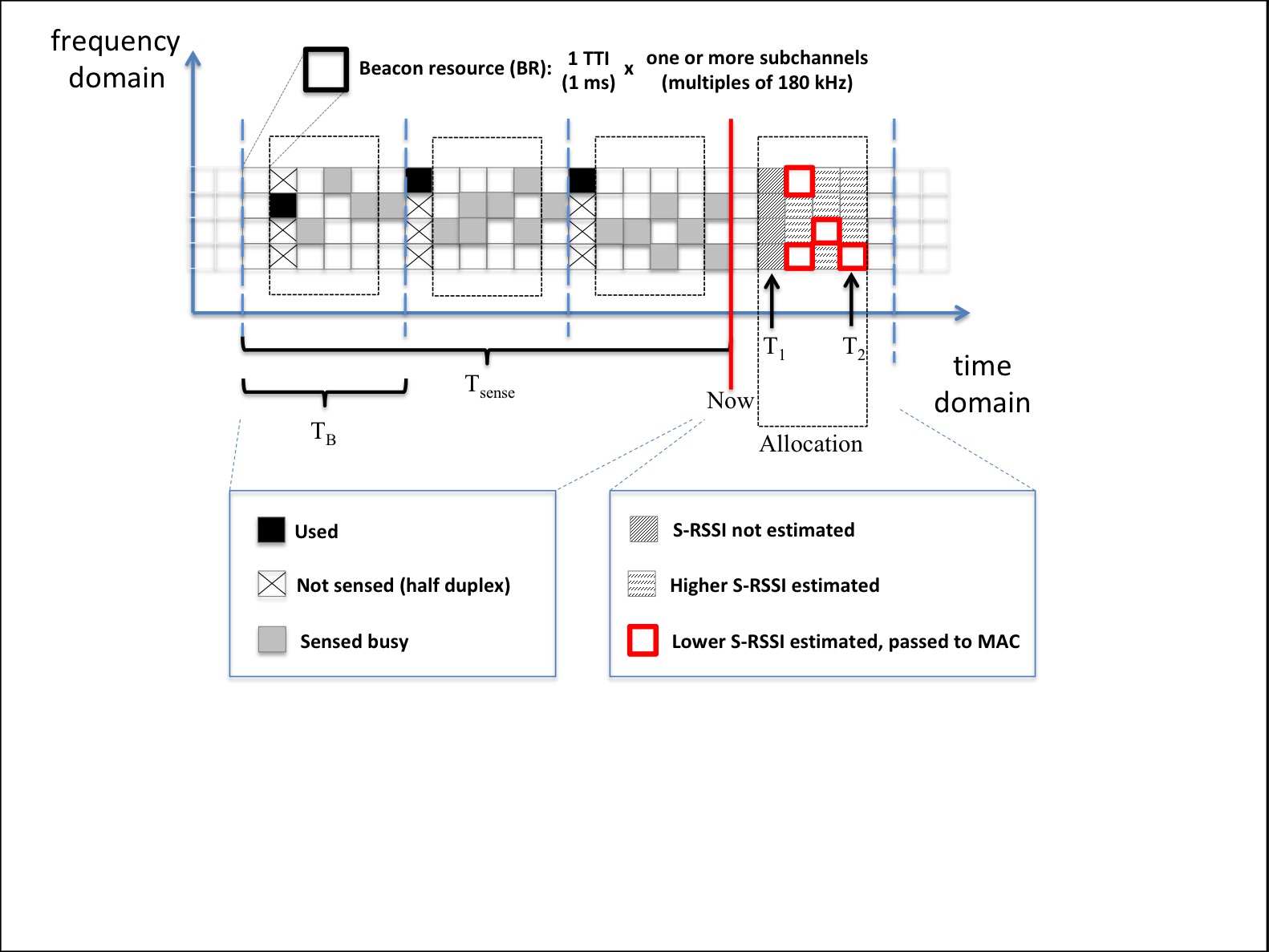}
	\caption{Example of BRs in the time and frequency domain. In the example, there are four BRs per subframe (in the frequency domain) and one beacon period $\Tbeacon$ lasts six TTIs, $\Tsense=3\;\Tbeacon$, $\Tone=2$, $\Ttwo=5$. It follows $\nRes=4 \cdot 6 = 24$, of which 16 are within $\Tone$ and $\Ttwo$. Assuming $\PercR=0.2$, the number of BRs passed to the MAC layer is $\nCandidates=4$.}
	\label{Fig:example}
\end{figure*}

The rest of the paper is organized as follows: in Section~\ref{sec:mode4}, the main characteristics of short-range C-V2X are summarized and the Mode~4 algorithm is briefly introduced; in Section~\ref{sec:modelling}, the models and settings adopted for numerical results are detailed; Sections~\ref{sec:PHY} and~\ref{sec:MAC} then focus on the impact of parameters at the PHY and MAC layer, respectively, followed by summary results and discussion in Section~\ref{sec:summary}; finally, in Section~\ref{sec:conclusion} we provide our conclusions.

\section{LTE-V2V and Mode 4 in Brief}\label{sec:mode4}



\setcounter{table}{0}
\input{TableParameters_values_clean.tex}

\subsection{LTE-V2V}

The concept of \ac{D2D} communications was initially introduced in Release~12 of LTE using the term \textit{sidelink} to differentiate from downlink and uplink. The new interface for this scope, called PC5, was explicitly designed for proximity services, and has been enhanced in Release~14 to address also vehicular scenarios, where the high mobility of nodes and the possibly high carrier frequency make the channel estimation more challenging.

The new technology, hereafter denoted as LTE-V2V to highlight the reference to the present release of the standards (i.e., LTE) and the case of car-to-car communication, is part of the \ac{C-V2X}, which promises to cover all use cases in a single chipset and to provide a continuous evolution with backward compatibility in the following releases.

Like LTE uplink, LTE-V2V adopts \ac{SC-FDMA} at the PHY and MAC layers, with the time-frequency domains organized into orthogonal resources called resource blocks. 
Resource blocks are allocated in pairs, corresponding to 180 kHz bandwidth (12 subcarriers with 15 kHz space) and 1~ms duration (14 OFDM symbols, of which 9 carry data, 4 are used for channel estimation, and one for timing adjustments and possible tx-rx switch).  The minimum allocation time interval is 1\;ms and is also denoted as \ac{TTI}.
In LTE-V2V, resource blocks are also grouped in the frequency domain into subchannels, which are all of a given size (set by the network, with the constraints detailed in \cite{3GPP_TS_36_331}). Each subchannel can carry at most one data packet, although one data packet can span over more than one subchannel. More specifically, each data packet, also known as \ac{TB}, has an associated control message, called \ac{SCI}, which requires 2 pairs of resource blocks. A \ac{TB} and the associated \ac{SCI} must be transmitted in the same subframe, but can be allocated on adjacent or non-adjacent resource blocks. In the former case (adjacent), subchannels carry both TBs and SCIs, with the SCI transmitted in the first allocated subchannel. In the latter case (non-adjacent), specific resources are reserved for \acp{SCI} and subchannels are only occupied by TBs. The number of subchannels allocated to carry a packet depends on the kind of allocation, the subchannel size, the TB size, and the adopted \ac{MCS}.

As mentioned, LTE-V2V has two possible approaches to allocate the resources for transmissions, namely Mode~3, where the network is in charge of performing the allocation and communicating it to the vehicles via signalling channels, and Mode~4, not requiring any intervention by the network. In both cases, a key traffic flow to be allocated is given by the cooperative awareness service, which means the broadcasting from each vehicle of periodic messages, hereafter called beacons\footnote{In this work we will use the generic term beacon	for the messages that broadcast the cooperative awareness information. Such beacons correspond, for example, to the \acp{CAM} of ETSI \cite{3GPP_EN_302_637_2} or a subclass of the \acp{BSM} of SAE \cite{SAE_DSRC_J2945_1}.}, detailing their status and movements. This service will play a key role in future \ac{CAV} networks for both safety applications and data routing, since it allows each node to have continuously an updated knowledge of its neighbourhood. Given the periodic nature of the transmissions, allocations are in such case performed on an \ac{SPS} basis, where the same subchannels are periodically reserved for some time in order to reduce the associated signalling.


\subsection{The 3GPP Mode 4 Algorithm in Brief}\label{subsec:Mode4}

The algorithm detailed by 3GPP for Mode 4 is divided into a PHY layer part \cite{3GPP_TS_36_213} and a MAC layer part \cite{3GPP_TS_36_321}. Hereafter, we provide a brief overview, whereas details on each of the mentioned parameters will be given later in Sections~\ref{sec:PHY} and~\ref{sec:MAC}. An example, including most of the described parameters, is shown in Fig.~\ref{Fig:example}.

Before entering in the description of the algorithm, we will introduce the concept of \ac{BR}, largely used in the further.

\subsubsection{Beacon resources (BRs)} The objective of the algorithm is to identify an appropriate group of subchannels 
to allocate one beacon, with the aim to maximize the probability of correct reception by neighbouring vehicles. Given the periodicity (a message every beacon period $\Tbeacon$) and adopting the SPS approach, once a beacon is allocated, the same subchannels are kept every $\Tbeacon$ for a given time. This implies that the selection process focuses on the next time window lasting $\Tbeacon$ and on all the groups of subchannels able to carry the beacon during that interval. Such groups are hereafter denoted as \acp{BR} and correspond to the single-subframe resources of 3GPP in \cite{3GPP_TS_36_213}. 

Given the size of the beacon, its generation periodicity, the used MCS, the size of subchannels, and the adjacent/non-adjacent allocation of SCIs, the generic node can calculate the number of messages that can be allocated in each subframe and create the time-frequency grid of \acp{BR} in one beacon period $\Tbeacon$. The number of \acp{BR} in one $\Tbeacon$ will be denoted as~$\nRes$. 
Please remark that the \acp{BR} are in principle orthogonal to each other (they do not interfere), except for some \ac{IBE} when they share the same TTI.

\subsubsection{PHY layer} At the PHY layer, the node continuously reads decodable \acp{SCI} and measures the average interference in each \ac{BR}, with the aim to estimate the occupation of the \acp{BR} in the next $\Tbeacon$. Measurements older than a given period $\Tsense$ are discarded, thus $\Tsense$ represents a sensing interval.

Given this information, the node focuses on the portion of BRs in the next $\Tbeacon$ that lay in an interval $\Tone$ to $\Ttwo$ \acp{TTI}, where $\Tone$ and $\Ttwo$ are parameters. Within this portion, the node considers as candidates only those that
\begin{enumerate}
	\item have been monitored; e.g., due to half duplex limitations, a node cannot sense during a \ac{TTI} if it transmits;
	\item are estimated as not used, either because known as not occupied by the associated \ac{SCI} or since the average measured \ac{RSRP} is below a given threshold $\Pthr$.
\end{enumerate}

The node then sorts the candidate \acp{BR} in terms of average \ac{S-RSSI} and selects the portion $\nCandidates$ with the lowest value, where $\nCandidates = \left\lceil \PercR \cdot \nRes \right\rceil$, $\PercR$ is a parameter, and $\lceil \cdot \rceil$ is the ceiling function.
If the number of candidates is smaller than $\nCandidates$, then $\Pthr$ is increased by 3~dB and the previous steps are repeated. Once the required number of \acp{BR} is reached, those selected are passed to the MAC layer.

\subsubsection{MAC layer} At the MAC layer, a  BR is randomly selected among the received $\nCandidates$. Given SPS, the BR is then reserved for a certain number of beacon periods, randomly selected within $\nMin$ and $\nMax$.

Once this time interval expires, a new resource allocation is performed with probability $1-\pkeep$. If a reallocation is commanded, the new BR is again randomly selected within the $\nCandidates$ received from the PHY layer. If a reallocation is not commanded (i.e., with probability $\pkeep$), the same BR is kept. Both if a new allocation is performed or not, a new random duration is set following the described rules and the process continues. 

\section{Modelling and Simulation Settings}\label{sec:modelling}

The results shown in the further have been obtained using the LTEV2Vsim simulator \cite{CecBazMasZan:C17}, which was designed for the investigation of resource allocation algorithms in C-V2X. In this section, a brief description of the main models and settings used for the simulations is provided.
The main settings are also summarized in Tables~\ref{Tab:parameters} and~\ref{Tab:Notations}, together with the values adopted if not differently specified.

\begin{figure} [t]
	\centering
	\includegraphics[width=0.9\linewidth,draft=false]{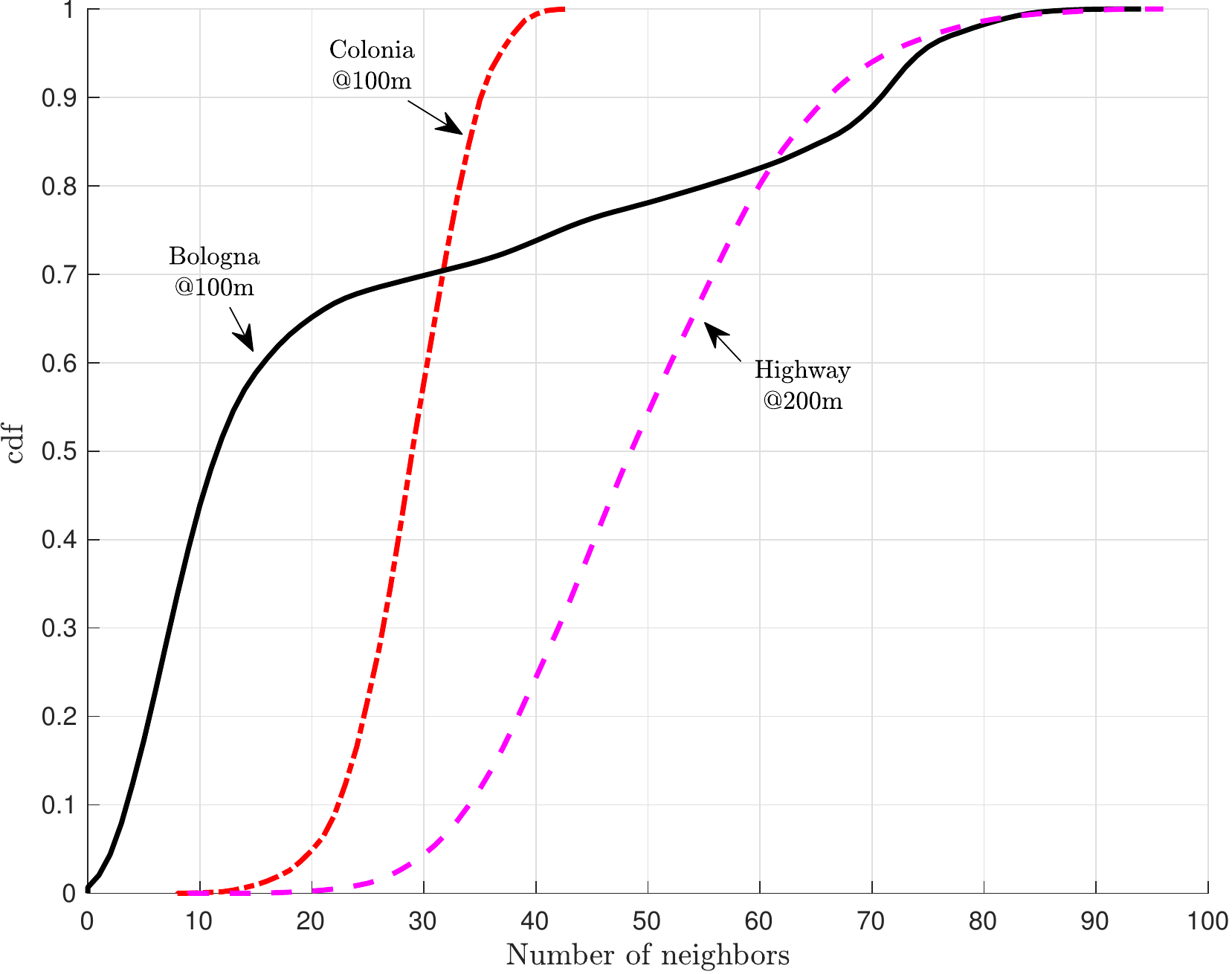}
	\caption{Cumulative distribution function of the number of neighbours in the various scenarios.}
	\label{Fig:neighbors}
\end{figure}

\subsection{Cooperative Awareness Settings and Definition of Neighbour}

To simplify the  scenario, we assume that all nodes have messages of the same size, generated at the same frequency, and transmitted using the same \ac{MCS}. Given these assumptions, the \ac{BR} grid and the number of beacon resources~$\nRes$ is the same for all vehicles. Please note that adopting the same size for all messages is equivalent to assume different sizes transmitted using the same number of resource blocks (as done for example in \cite{NguShaSudKapEtc:C17}) and focus the performance investigation on the largest one as the worst case.\footnote{For example, 3GPP suggests in \cite{3GPP_TR_36_885} sequences of four \acp{CAM} of 190~bytes and one of 300~bytes. Our results still hold if the smaller messages use the same amount of resources than the larger ones, adopting a lower coding rate. A different option is to use less resources for the smaller packets, which however causes high inefficiencies with Mode~4, as explained in \cite{MolGoz:J17}.}

Beacons are assumed of size $\bBytes$, with periodicity $\fBeacon$. Equivalently, beacons are periodically generated with a constant periodicity $\Tbeacon=1/\fBeacon$. In particular, $\fBeacon$ is set to 10~Hz, which is the value most commonly adopted and $\bBytes=300$~bytes, which is the largest size suggested by 3GPP in \cite{3GPP_TR_36_885}.  


Beacons are broadcast, thus each of the other vehicles is a potential receiver. However, the importance of a message reduces with the distance. For this reason, we focus on a given maximum distance, set to 100\;m in the urban and 200\;m in the highway scenarios, and we denote as \textbf{neighbours} all the vehicles within such range from the source. 

\subsection{Output Metrics}

The following output metrics will be used.
\begin{itemize}
	\item \textbf{\ac{PRR}}: the average ratio between the number of neighbours correctly decoding a beacon and the total number of neighbours;
	\item \textbf{\ac{UD}}: given a destination and source couple, it is the time difference between the instant a message is correctly received and the instant the last of the previous messages was correctly received. The UD quantifies how long a node does not receive any update from one neighbour and implicitly gives information about the correlation among errors.
\end{itemize} 

\input{TableSettings.tex}

\subsection{Scenarios}\label{subsec:scenario}

Results are provided in the following three scenarios.
\begin{itemize}
	\item \textbf{Cologne - Urban, medium density}: The scenario is a 1.85$\times$1.85~km$^2$ portion, at 7:30, of the urban trace detailed in \cite{UppTruFioBar:J14}; on average, there are 925 vehicles, each with 14.8 neighbours within 100~m (standard deviation 8.8), thus the network is moderately dense;
	\item \textbf{Bologna - Urban, congested}: The scenario is a 1.6$\times$1.3~km$^2$ urban area, denoted in \cite{BazMasZanCal:J16} as congested; on average, there are 667 vehicles, each with 25.4 neighbours within 100~m (standard deviation 25.4); certain roads are highly loaded, with even long tailbacks at some junctions;
	\item \textbf{Highway - High density}: The scenario, detailed in \cite{BazMasZanCal:J16}, corresponds to approximately 16\;km of a 3+3 lanes highway; on average, there are 2015 vehicles with 49.4 neighbours within 200~m (standard deviation 12.5); the road is highly loaded.
\end{itemize}

In Fig.~\ref{Fig:neighbors}, the distribution of the number of neighbours that the generic vehicle has in each scenario is shown, considering a maximum distance of 100~m in the urban scenarios (Bologna and Cologne) and 200~m in the Highway scenario. As observable, they have different characteristics and allow to evaluate the performance under various densities. Please note that the upper 20\% of vehicles in Bologna and in the highway scenario have a similar number of neighbours, even if the considered range is the half; this implies that in Bologna there is a relevant portion of nodes that are subject to heavy interference conditions. In addition, whereas all links are in \ac{LOS} conditions in the Highway scenario, communications are often affected by \ac{NLOS} conditions in the urban cases. 

\subsection{Propagation and Settings at the PHY layer}

\begin{figure} [t]
	\centering
	\includegraphics[width=0.9\linewidth,draft=false]{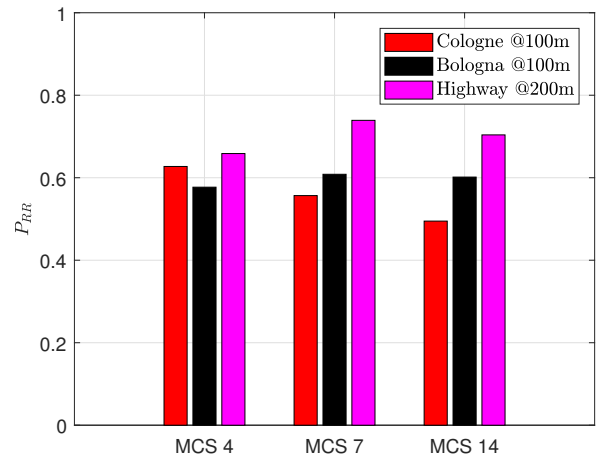}
	\caption{Packet reception ratio varying the scenario and MCS, assuming the settings of Mode~4 detailed in Table~\ref{Tab:parameters}.}
	\label{Fig:MCS}
\end{figure}

All devices are assumed transmitting with 23~dBm power and using antennas with 3~dB gain. As for the propagation, the WINNER+ model, scenario B1, is adopted as recommended by 3GPP in \cite{3GPP_TR_36_885}. The model also includes log-normal correlated shadowing, with variance 3~dB in \ac{LOS} and 4~dB in \ac{NLOS} and with decorrelation distance 10~m in urban and 25~m in highway scenarios. A packet is assumed to be correctly received if the corresponding \ac{SINR} is larger than a minimum threshold. All details about the calculations, which take into account co-channel interference and \ac{IBE}, are provided in Appendix A.

The typical channel bandwidth of 10~MHz is assumed, which corresponds to 50~pairs of resource blocks per subframe. A non-adjacent allocation of SCIs, with four subchannels of 10~pairs (and the remaining used by the SCIs) are assumed. 

Preliminary simulations have been then performed to set the \ac{MCS} per each scenario. With the given assumptions, the options are to either allocate in each subframe one beacon occupying four subchannels with MCS 4, or two beacons occupying two subchannels each with MCS 7, or four beacons, one per subchannel, with MCS 14 (all the other \acp{MCS} would reduce the reliability without reducing the occupation of subchannels). 

In Fig.~\ref{Fig:MCS}, the PRR is shown for \ac{MCS}~4, 7, and~14 in the three scenarios. In general, the PRR is higher in the Highway scenario since there are no buildings impairing the communication and all links are in \ac{LOS}. Comparing the various \acp{MCS}, it can be noted that the results are the consequence of a trade-off: on the one hand, a higher value implies more available \acp{BR}, thus a lower collision probability, while on the other, increasing the \ac{MCS} raises the required minimum \ac{SINR}, thus causing a higher probability that the received power is insufficient. As observable, the best performance is provided by \ac{MCS}~4 in the lightly loaded scenario of Cologne, where there are on average less competing nodes, and by \ac{MCS}~7 in the other two. For this reason, these \acp{MCS} are adopted in the further and it follows that $\nRes=100$ in the Cologne scenario and $\nRes=200$ in the Bologna and Highway scenarios. 

\subsection{Preliminary Considerations: Hidden Node Probability}

Before discussing the impact of the various parameters, it is interesting to estimate how frequent is the event that an interfering signal is not revealed by a node, thus making the sensing procedure ineffective. To this aim, we calculate the probability that an interferer is not sensed, hereafter called hidden node probability, as detailed in Appendix~B.

\begin{figure}
	\centering
	\includegraphics[width=0.9\linewidth,draft=false]{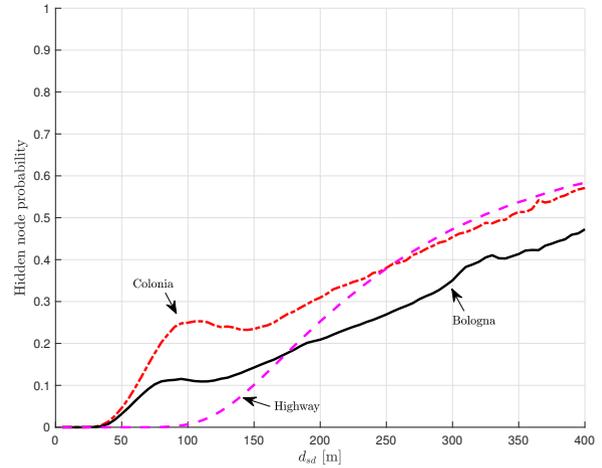}
	\caption{Hidden node probability in the three scenarios.}
	\label{Fig:hiddenNodes}
\end{figure}

The hidden node probability varying the source-destination distance $\dist$ is shown for the three scenarios in Fig.~\ref{Fig:hiddenNodes}. As expected, the hidden node probability is negligible when the source is near to the destination and then increases when the distance rises. Some differences can be observed in the three scenarios, due to the peculiar road topologies. In particular, especially for short distances, the absolute values vary among all scenarios and a non-monotonic behaviour can be noted in Cologne and Bologna. This is a direct consequence of the LOS/NLOS conditions, that strictly depend on the length of the road segments and the number of intersections.

As the main derivation from Fig.~\ref{Fig:hiddenNodes}, it is important to note that the hidden node probability remains limited within reasonable distances. In particular, it results below 10\% up to 50~m in the urban scenarios and almost 150~m in the Highway and do not exceed 30\% as a worst case within the distances considered hereafter (i.e., 100~m in urban, 200~m in Highway). This implies that the sensing procedures are indeed potentially effective against most interferers.

\begin{figure*}[t!]
	\centering
	\subfigure[Varying the sensing period.]
	{\includegraphics[width=0.45\linewidth,draft=false]{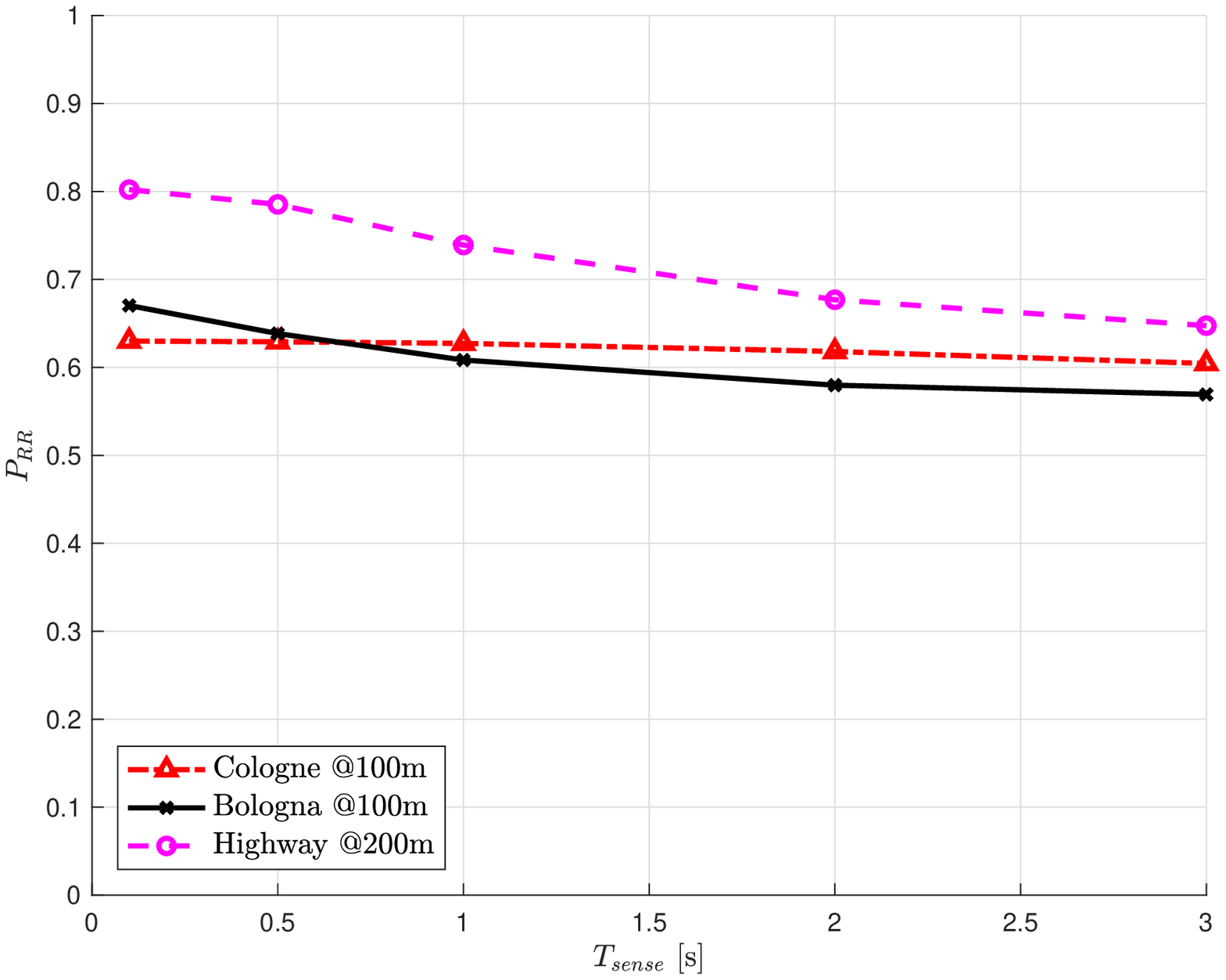}\label{fig:varyTsense}}~~~
	\subfigure[Varying the power threshold.]
	{\includegraphics[width=0.45\linewidth,draft=false]{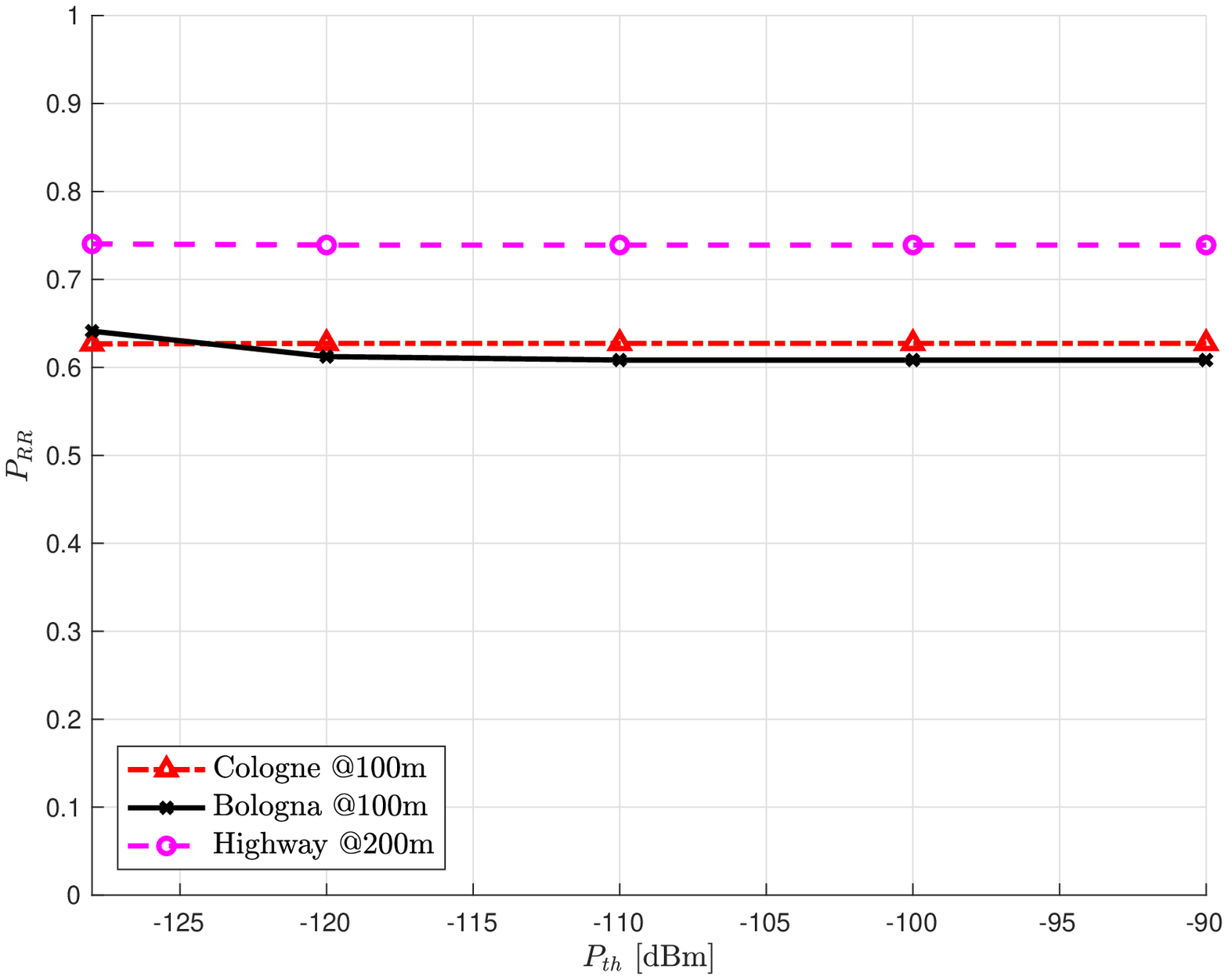}\label{fig:varyPthr}}\\
	\subfigure[Varying the portion of resources.]
	{\includegraphics[width=0.45\linewidth,draft=false]{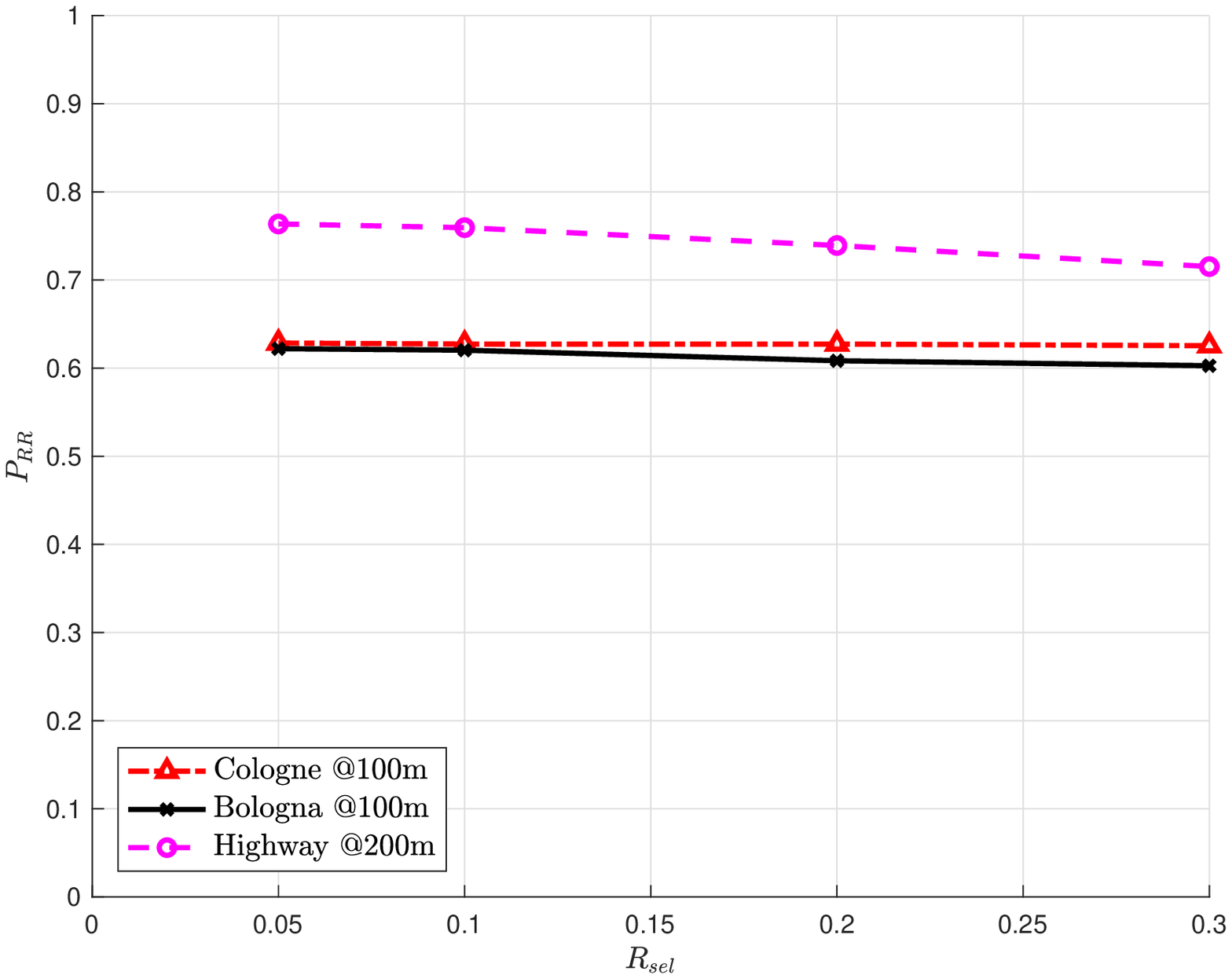}\label{fig:varyResources}}~~~
	\subfigure[Varying the time window.]
	{\includegraphics[width=0.45\linewidth,draft=false]{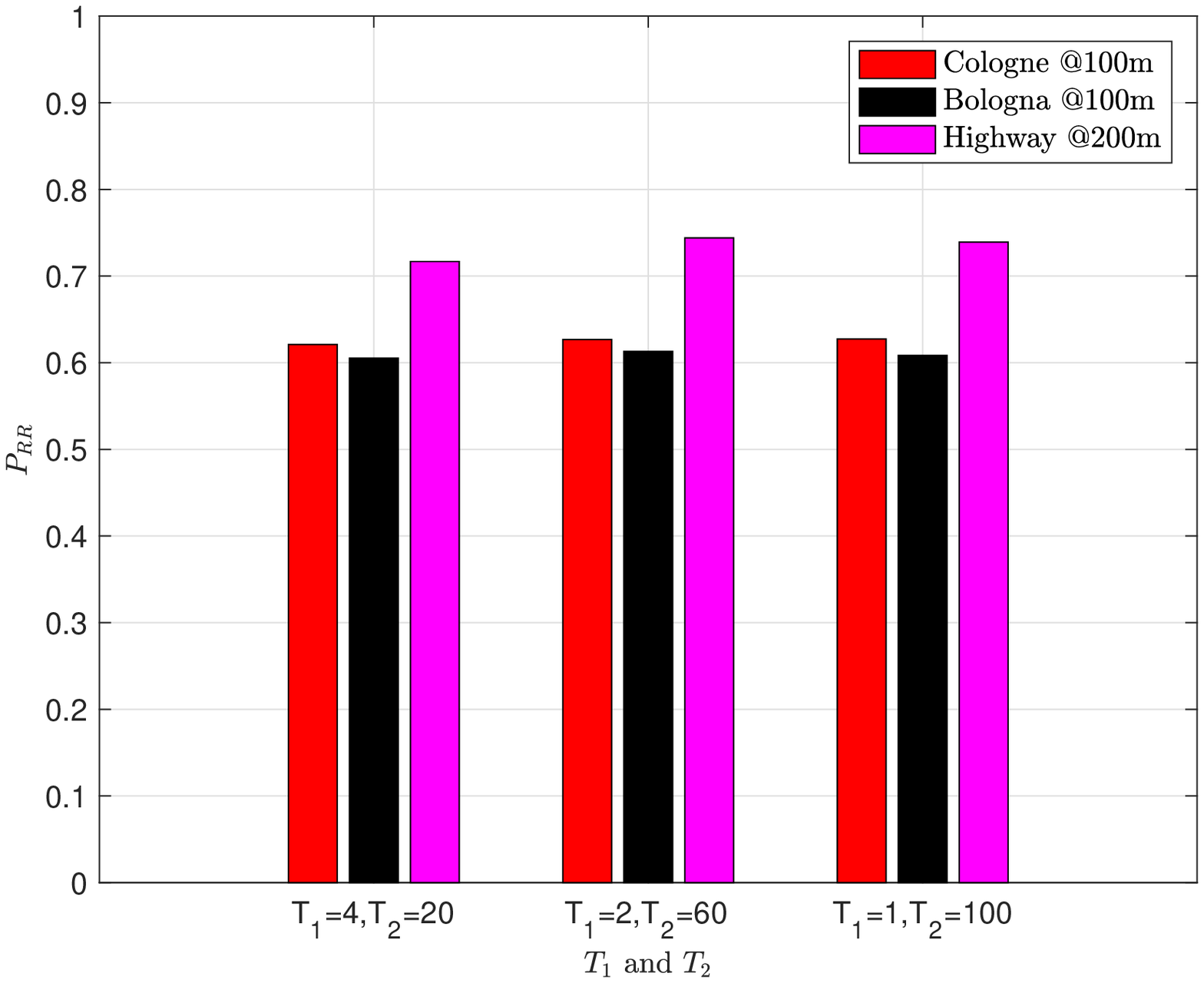}\label{fig:varyT1T2}}
	\caption{Impact of PHY parameters: simulations in  realistic scenarios. The values listed in Table~\ref{Tab:parameters} are used for those settings not explicitly indicated.}\label{fig:varyPHY}
\end{figure*}

\section{Impact of Mode 4 PHY settings}\label{sec:PHY}

As discussed in Section~\ref{subsec:Mode4} and summarized in Table~\ref{Tab:parameters}, the algorithm is characterized by several parameters at both PHY and MAC layers. In this section, the possible values of those at the PHY layer and their impact will be detailed. Since PRR and UD lead to the same conclusions, curves are shown only in terms of PRR to limit the number of figures; comments to UD are provided only when relevant.

\subsubsection{Sensing period} $\Tsense$ is the time interval during which the node decodes the \acp{SCI} and measures the average interference power per each \ac{BR}.  Since the nodes are mobile and reallocations are continuously performed, averaging over long periods increases the risk of inaccurate or outdated measurements. In principle, the lower it is and the better is the estimation for the next future. At the same time, however, it must be long enough to individuate at least one transmission from all neighbouring vehicles, i.e., it should be longer than their beacon period. In the standard, the value is fixed by specifications, depending on the duplexing type, and normally corresponds to 1\;s. Please note that 1\;s allows to sense at least one transmission from any node having a beacon periodicity of 1~Hz, which should be the lowest value under normal operation conditions.

The impact of a variation of $\Tsense$ in the investigated scenarios is provided in Fig.~\ref{fig:varyTsense}. As expected, except for Cologne, where the sparsity of the network makes results negligibly dependent on $\Tsense$, \ac{PRR} reduces increasing $\Tsense$. What is interesting to notice is that a high impact on PRR is observable when $\Tsense$ gets lower than 1~s (up to 10\% higher PRR). As explained previously, this is indeed coherent with the presence of a beacon period of 100~ms. For the same reasons and given that higher beacon periods are possible, simply reducing $\Tsense$ below 1~s is not a viable solution in general. However, the significance of the improvement suggests that some effort could be posed to enhance the algorithm, for example by either making it variable with the settings of the neighbours or adding mechanisms that better identify and discard the outdated information.

\subsubsection{Power threshold} $\Pthr$ defines a power level below which a BR is assumed as available, independently on what inferred from the decoded \ac{SCI}. This allows to control the interference level that is considered acceptable and make the selection process more or less stringent. $\Pthr$ is set by the upper layers, depending on the priority $a$ of the transmitter and $b$ of the receiver (both within 0 and 7). Specifically, it is set to a value in the range $[-128,-2]$~dBm following the formula
\begin{equation}
\Pthr = -128 + 2 \left( a \cdot 8 + b \right)\;\text{dBm}\;.
\end{equation}

\begin{figure*}[t]
	\centering
	\subfigure[Statistic of the duration before reallocation.]
	{\includegraphics[width=0.45\linewidth,draft=false]{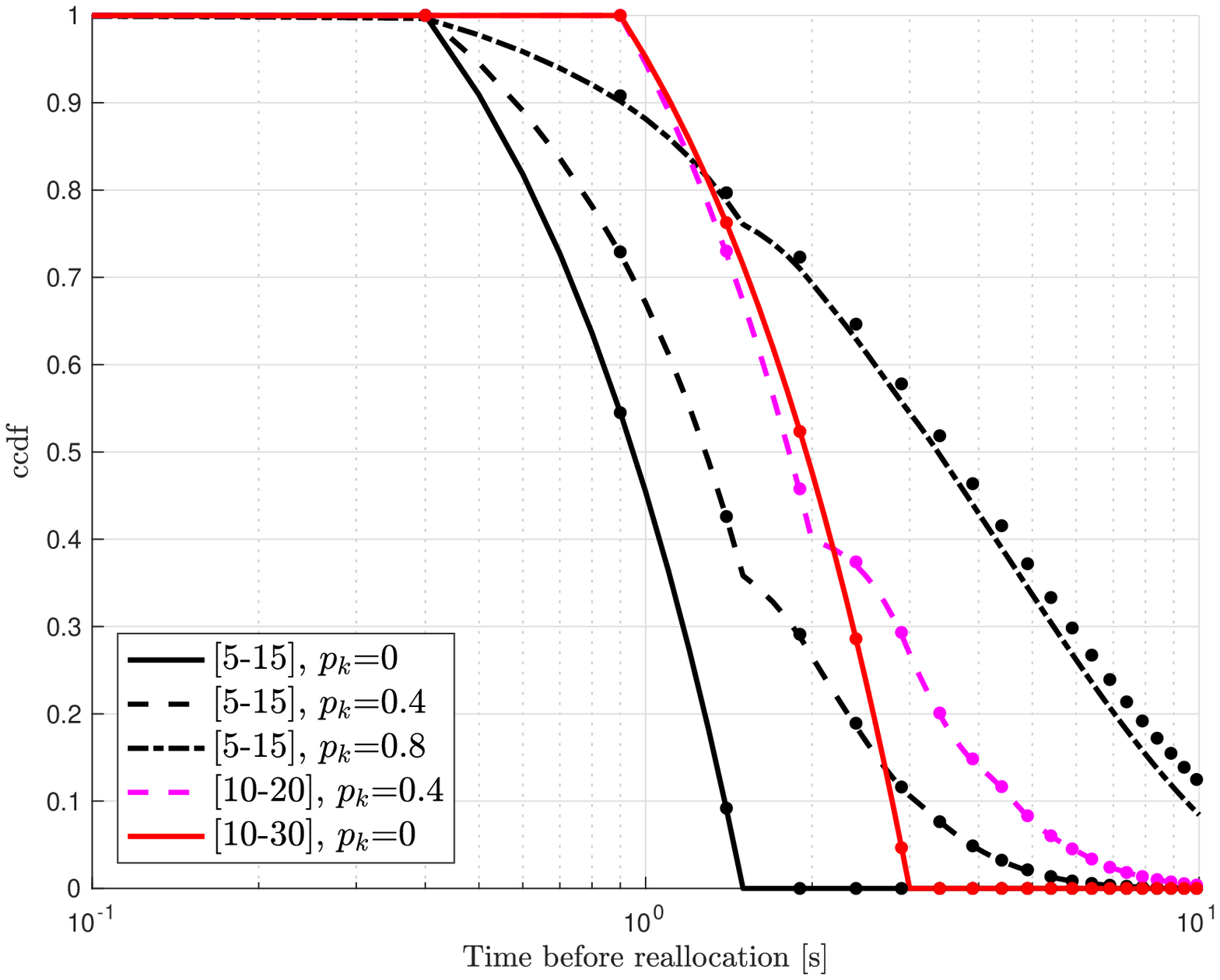}\label{fig:analysisKeep}}\;\;
	\subfigure[Probability of reallocation during the sensing period.]
	{\includegraphics[width=0.45\linewidth,draft=false]{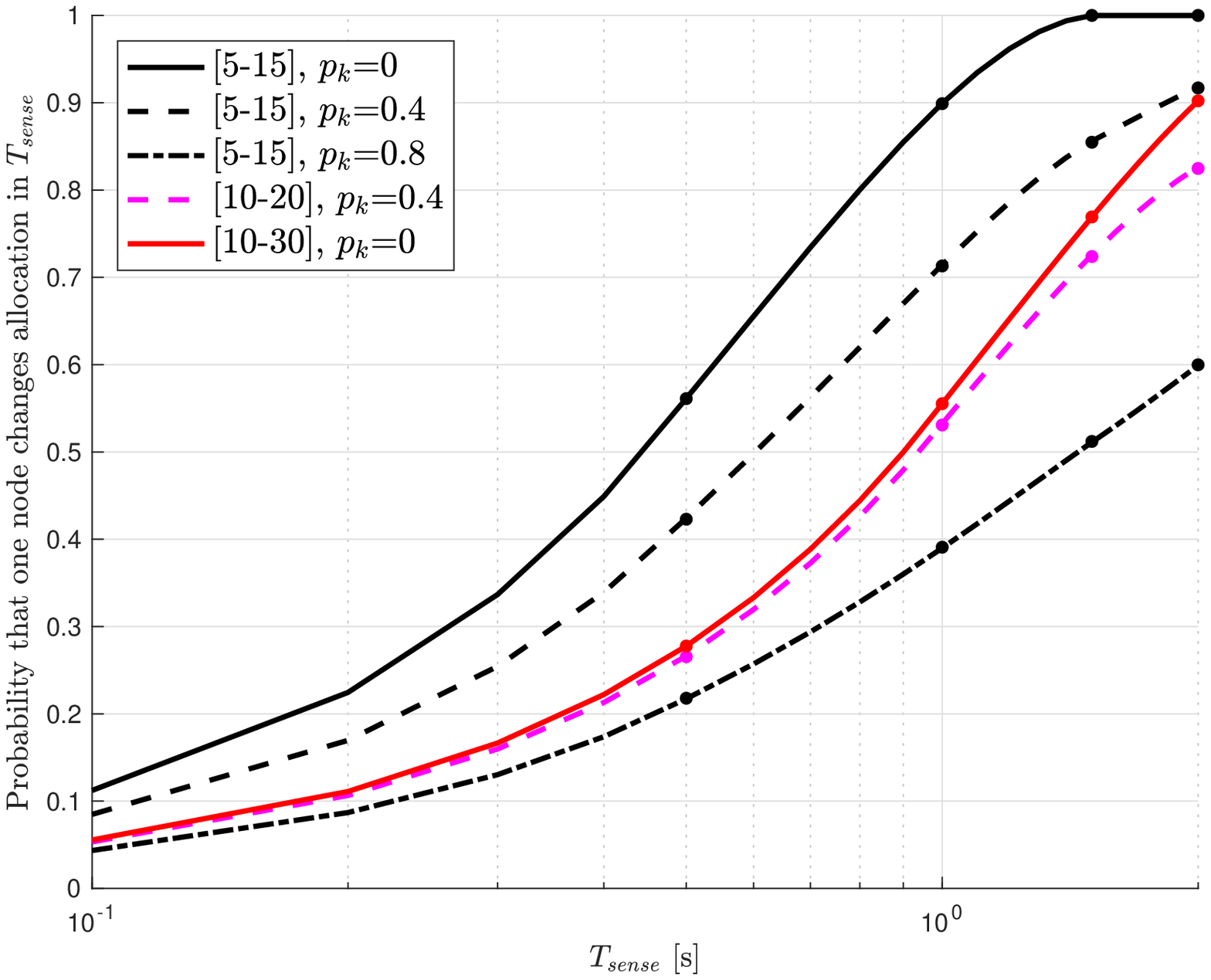}\label{fig:analysisTsense}}
	\caption{Impact of MAC parameters: analysis in simplified scenarios. Results are shown for various combinations of [$\nMin$, $\nMax$], and $\pkeep$. Lines are obtained with \eqref{eq:PTkeep} and \eqref{eq:probChange}, whereas dots correspond to simulations used to validate the analysis.}\label{fig:analysisMAC}
\end{figure*}

The impact of $\Pthr$ is shown in Fig.~\ref{fig:varyPthr}. As observable, the impact is irrelevant in both Cologne and Highway. Only in Bologna, a small $\Pthr$ is shown to improve the performance of about 5\% compared to a high $\Pthr$. In this case, there are a few congested intersections, where vehicles have more than 70 neighbours in a range of 100~m. To better understand these results, please recall that only a portion of the resources assumed available will be then passed to the MAC, starting from those that have less interference. This implies that $\Pthr$ is relevant only if a very large number of resources is affected by an interference higher than $\Pthr$, thus in very dense scenarios. Summarizing, the lower is $\Pthr$ and the higher is the \ac{PRR}, although some impact is only observable in crowded conditions. 

\subsubsection{Portion of resources} The parameter $\PercR$ is then set to control the number of \acp{BR} passed to the MAC layer. $\PercR$ is fixed to 0.2 by specifications.

Apparently, the lower is $\PercR$, the higher is the probability to select a resource with negligible interference, thus a smaller $\PercR$ might be expected to perform better. Indeed, this effect can be observed in Fig.~\ref{fig:varyResources}, even if the variation of PRR is very small:  if we look at Bologna and Highway, a slight decrease while increasing $\PercR$ can be noted, especially with $\PercR \geq 0.1$. However, with simulations not shown here for brevity, it was noted that a too low $\PercR$ causes a slight increase in some cases.\footnote{For example, this happened assuming Bologna with MCS 4. In such case, the congested situation due to a very high density of vehicles is emphasized by few resources available ($\nRes=100$).} With a smaller set of resources for the random selection of the BR, in fact, the risk of a collision increases if the choice is performed by two nodes at the same time from the same pool. And having the same pool is rather frequent for nodes that are located near to each other and thus sense similar interference. This condition becomes anyway relevant only when the number of neighbours is very high compared to the number of resources. To summarize, the value 0.2 given by the standard appears as an acceptable compromise, even if 0.1 brought us to a small improvement (2-3\% in our experiments).

\subsubsection{Time window} $\Tone$ and $\Ttwo$  are two further parameters at the PHY layer that allow to restrict the interval of the allocation. The former, which must be between 1 and 4, indicates the first TTI where the allocation can be performed and gives a time margin to the device for the selection process. The latter, between 20 and 100, sets the last possible TTI and is used in the case of stringent delay requirements. 

The results, shown in Fig.~\ref{fig:varyT1T2} for three combinations of $\Tone$ and $\Ttwo$, show a negligible difference in terms of PRR. In terms of UD we noted a very slight increase 
only with $\Tone=4$ and $\Ttwo=20$, confirming the intuitive conclusion that a larger time window better randomizes the resource selection among the various users. The value of $\Tone$ and $\Ttwo$ should thus be reasonably set to maximize the window, once the constrains on processing and delay are applied.

\begin{figure*}[t]
	\centering
	\subfigure[Packet reception ratio.]
	{\includegraphics[width=0.45\linewidth,draft=false]{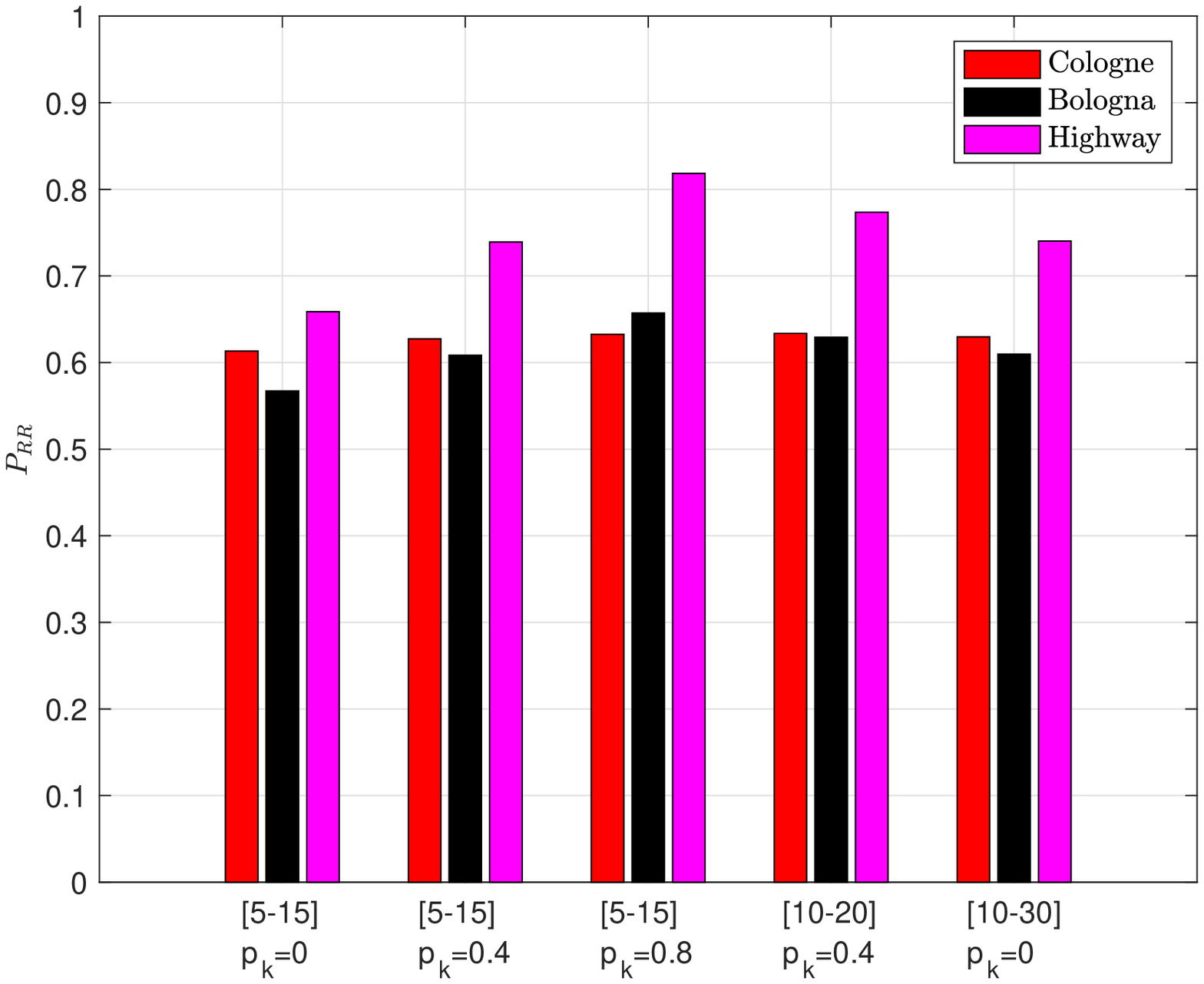}\label{fig:prrMAC}}\;\;
	\subfigure[99.99\% of the update delay.]
	{\includegraphics[width=0.45\linewidth,draft=false]{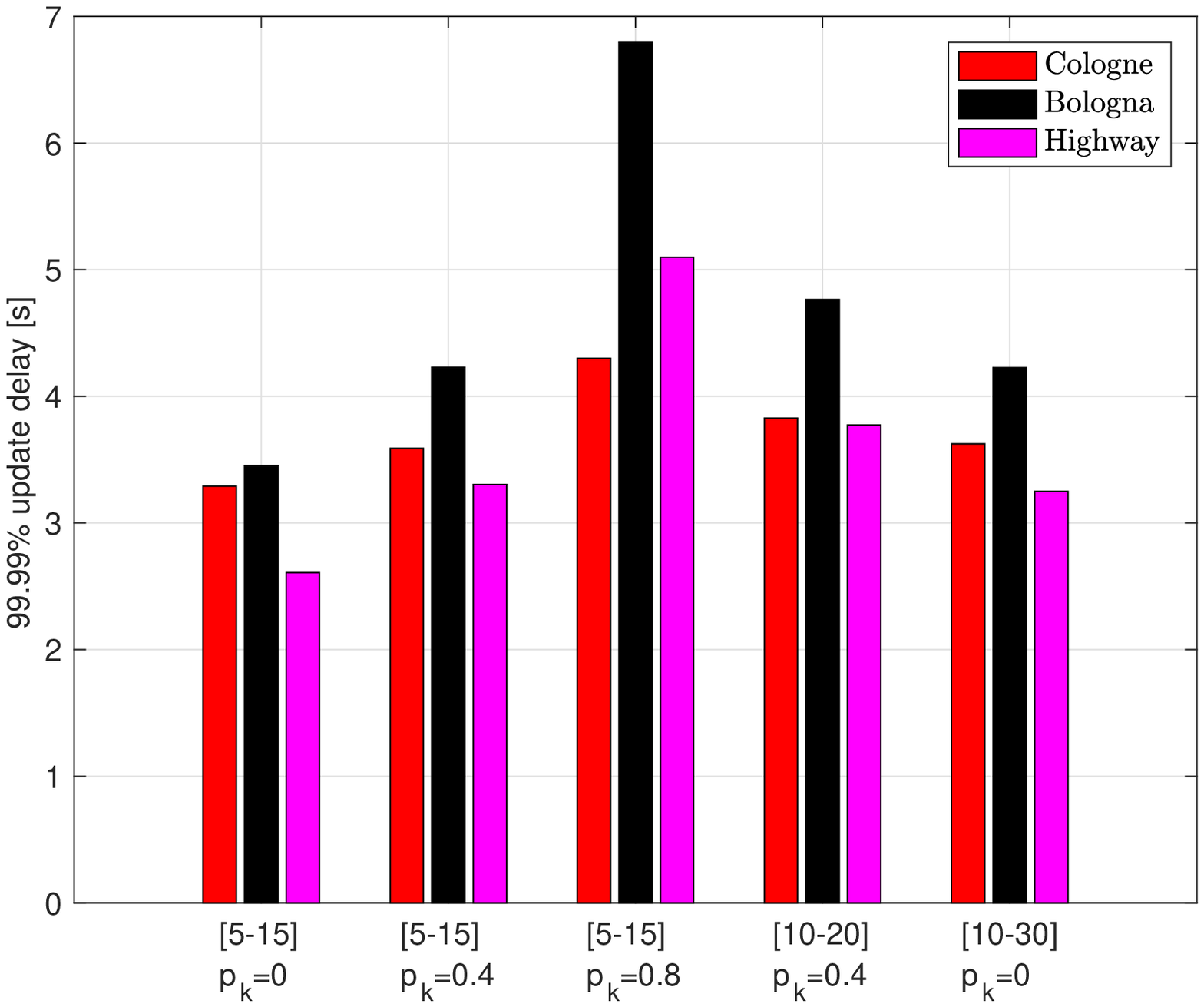}\label{fig:UDMAC}}
	\caption{Impact of MAC parameters: simulations in  realistic scenarios. Results are shown for various combinations of [$\nMin$, $\nMax$], and $\pkeep$. The values listed in Table~\ref{Tab:parameters} are used for all the other settings.}\label{fig:simsMAC}
\end{figure*}

\section{Impact of Mode 4 MAC settings}\label{sec:MAC}

In this section, the attention is moved to the MAC layer. The main parameters, summarized in Table~\ref{Tab:parameters}, are:
\begin{itemize}
	\item \textit{Time before evaluation.} Parameters $\nMin$ and $\nMax$ define the minimum and maximum number of beacon periods before a reallocation is considered (only considered and not always performed, as later clarified). Such an interval will be hereafter denoted as \textit{\ac{TBE}}. The actual duration is randomly selected with uniform distribution within $\nMin$ and $\nMax$. By standard, they are respectively fixed to 5 and 15 for a beacon periodicity of 10~Hz (higher values are defined for smaller frequencies); this means that the same RB is allocated for a duration that ranges between 0.5 and 1.5 seconds;
	\item \textit{Keep probability.} Once the selected number of beacon periods has expired, a new allocation is performed with probability $1-\pkeep$. Equivalently, with keep probability $\pkeep$, the same allocation is maintained for another random number of beacon periods. The value of $\pkeep$ can be chosen between 0 and 0.8.
\end{itemize}

Parameters $\nMin$, $\nMax$, and $\pkeep$ together determine the distribution of the duration of an \ac{SPS} allocation before a different scheduling is performed, hereafter denoted as \textit{\ac{TBC}}. Directly related to $\pkeep$, one \ac{TBC} is composed of a variable number of \ac{TBE}; each \ac{TBE} is in turn of variable duration, which depends on $\nMin$ and $\nMax$. 

It must be noted that an increase of \ac{TBC} has two opposite effects: on the one hand, it makes the use of the channel more stable, thus making the sensing from neighbouring nodes less affected by outdated information (recall the discussion on $\Tsense$); on the other hand, possible wrong selections are maintained for more time (for example, if two nearby nodes select the same resource they remain invisible for longer).

These two opposed effects are hereafter better highlighted, first focusing on specific metrics in a simplified scenario, and then showing PRR and UD in the realistic ones. With the sake of focusing on relevant cases, we compare five representative combinations: three with $\nMin=5$ and $\nMax=15$, varying $\pkeep$ from 0 to 0.4 and 0.8; the other two, assuming $\nMin=10$, $\nMax=20$, $\pkeep=0.4$ or $\nMin=10$, $\nMax=30$, $\pkeep=0$.

\subsection{Focus on Specific Metrics.} 
Let us first focus on the average duration before a reallocation is performed.
Following the algorithm at the MAC layer, the probability that the duration of one \ac{TBE} is equal to a number $n > 0$ of beacon intervals $\Tbeacon$ is
\begin{equation}
\Pnsingle(n)= \left \{
\begin{array}{ll}
1/n & n \in [\nMin,\nMax]\\
0 & \text{otherwise} \\
\end{array} 
\right.
\end{equation}
which can be re-written in the form of vector $\pvector \triangleq \left(\Pnsingle(1), \Pnsingle(2), ...\right)$.

As a consequence, the probability distribution $\Pntot(n)$ that the same allocation is kept for a given number $n$ of beacon periods is then obtained as 
\begin{equation}\label{eq:PTkeep}
\Pntot(n) = \left( 1 - \pkeep \right) \sum_{i=1}^{\infty} \pkeep^{i-1} \InvFouD{ \left(\FouD{\pvector}\right)^i  }
\end{equation}
where $\FouD{\cdot}$ denotes the discrete Fourier transform and the power raised to the vector denotes its application to each element. The demonstration is in Appendix C.

$\Pntot(n)$ quantifies the duration before reallocation and should not be too large to avoid that unfortunate allocations last for too long. With trivial elaborations from $\Pntot(n)$, 
in Fig.~\ref{fig:analysisKeep} it is shown the \ac{ccdf} of the time before reallocation. Simulation results that validate the analysis are also shown. As observable, if $\nMin=5$, $\nMax=15$, and $\pkeep=0.8$, the probability that the same resource is maintained for more than 10 seconds is approximately 0.1. This could be extremely dangerous, since it implies that two neighbouring vehicles simultaneously selecting a resource of the same \ac{TTI} (recalling the half duplex limitation) will remain hidden to each other for more than 10~s in 1 case every 100.

Looking now at the negative effect of a short time before reallocation, from \eqref{eq:PTkeep} we also derive the probability to observe a change during the sensing interval, which causes a wrong view of the occupied resources. To this aim, the probability that a change is performed within a given interval of $n^*$ beacon periods $\pCamb(n^*)$ is obtained as 
\begin{equation}\label{eq:probChange}
\pCamb(n^*) = 1 - \sum_{n = n^*}^{\infty} \frac{n - n^*}{n} \Pntot(n) \;.
\end{equation}
The proof is in Appendix D.

Assuming $\Tsense$ is a multiple of $\Tbeacon$, the probability that a change is performed during the sensing interval is $\pCamb(\Tsense/\Tbeacon)$, which should be obviously as close as possible to zero: in fact, all measurements performed during $\Tsense$ before the change alter the correct view of the interference. Results varying $\Tsense$ are shown in Fig.~\ref{fig:analysisTsense}. Simulations  that validate the analysis are also shown.  As observable, if the standard $[5,15]$ is assumed with  $\pkeep=0$, the probability to have a change during $\Tsense=1$~s is 0.9, which means that 90\% of the sensed nodes have changed their allocation during the sensing interval. This very high probability reflects on an inaccurate estimation of the interference and thus a reduced efficiency of the sensing process.

\subsection{Simulations in the Investigated Scenarios.} Simulation results in the considered scenarios are then shown in terms of PRR and 99.99\% of UD in Fig.~\ref{fig:simsMAC}. 

As observable, again the impact of different combinations of $\nMin$, $\nMax$, and $\pkeep$ have opposed effects on PRR and UD. Whereas increasing any of the values have a positive impact on PRR due to an higher stability on the resource usage and thus a more efficient sensing process, the impact is negative on the update delay, because the duration of wrong allocations is statistically longer. This effect makes plain that the optimal definition of the parameters at MAC layer is subject to an unavoidable trade-off.

Another interesting point is that acting on the window does not lead to significant improvements. This means that the modification of $\pkeep$ is enough to control the trade-off between PRR and UD and there is no necessity to modify $\nMin$ and $\nMax$.

\section{Summary results and discussion}\label{sec:summary}

In Fig.~\ref{Fig:summaryPrr} and Fig.~\ref{Fig:summaryUD}, the \ac{PRR} and \ac{UD} are respectively shown for all the addressed scenarios. The former shows the PRR varying the source-destination distance $\dist$, while the latter provides the \ac{UD} correspondent to a target percentile, varying such target. In each sub-figure, five curves are compared, corresponding to the following allocations:
\begin{itemize}
	\item IEEE 802.11p (with CSMA/CA, hidden terminals, capture effect, and so on), adopting QPSK with 1/2 coding rate, which corresponds to 6~Mbps raw data rate and is the \ac{MCS} normally used by default;
	\item Random, meaning that each vehicle changes allocation every $\Tbeacon$, selecting one of the $\nRes$ \acp{BR} at random;
	\item Standard protocol, with $\pkeep=0$ (the minimum) and the other parameters set as in Table~\ref{Tab:parameters};
	\item Standard protocol, with $\pkeep=0.8$ (the maximum) and the other parameters set as in Table~\ref{Tab:parameters};
	\item An optimized Mode 4 where $\pkeep=0.8$, $\Pthr=-128$~dBm, and some parameters are changed outside the specifications, with particular reference to $\Tsense=0.1$~s and $\PercR=0.1$.
\end{itemize}
IEEE 802.11p is considered as a benchmark technology for this application. The random allocation is also used as benchmark, since it is the simplest way to allocate resources in LTE-V2V. 
Then, in addition to considering the standard protocol with the two extremes of $\pkeep$, the last item corresponds to the maximum \ac{PRR} obtainable with the detailed algorithm by modifying all parameters. 

From Fig.~\ref{Fig:summaryPrr} and Fig.~\ref{Fig:summaryUD}, the following observations can be derived.

\setcounter{subsubsection}{0}
\subsubsection{LTE-V2V vs. IEEE~802.11p} 
\begin{figure}[t!]
	\centering
	\subfigure[Cologne.]
	{\includegraphics[width=0.88\linewidth,draft=false]{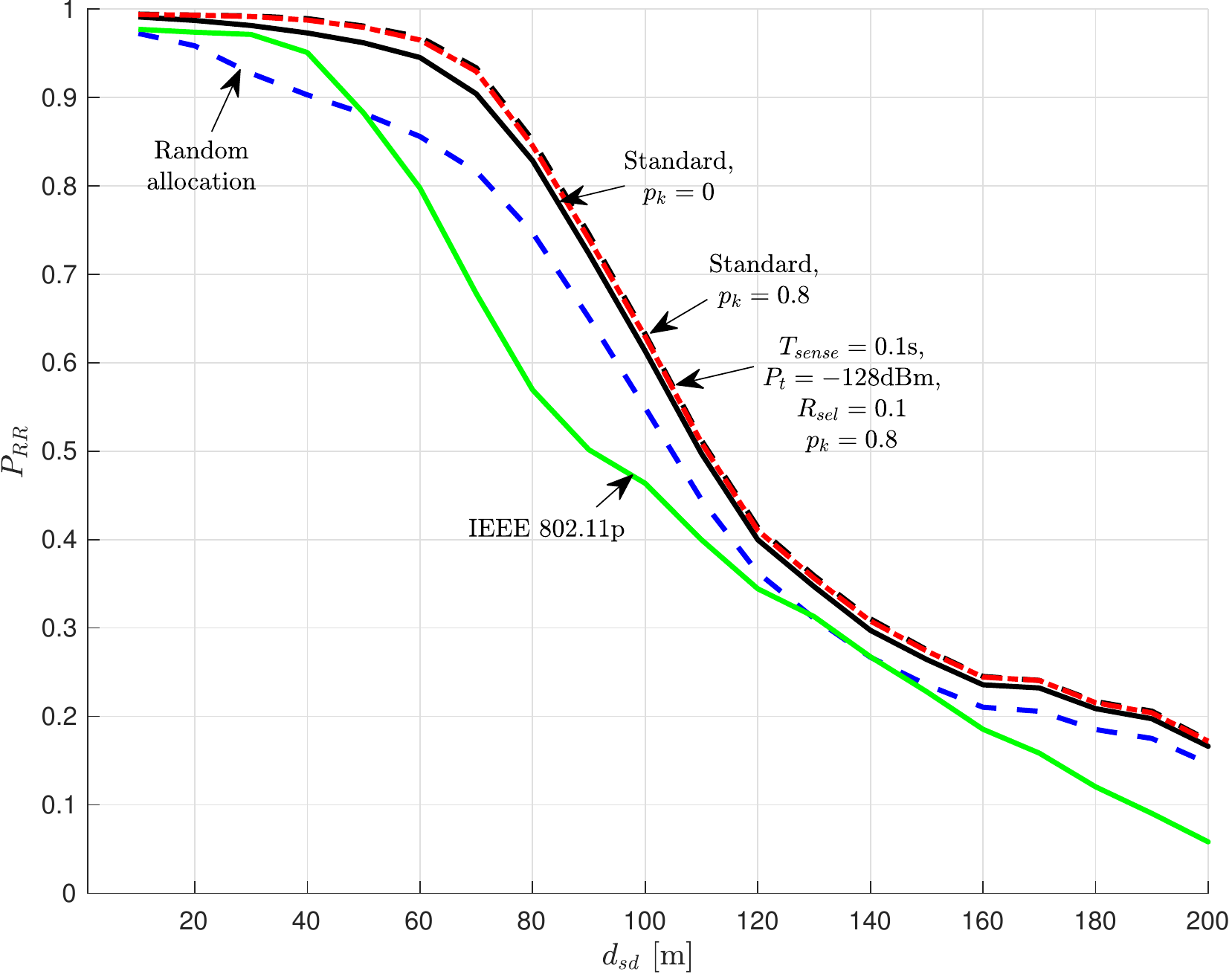}\label{fig:summaryPrrColonia}}
	\subfigure[Bologna.]
	{\includegraphics[width=0.88\linewidth,draft=false]{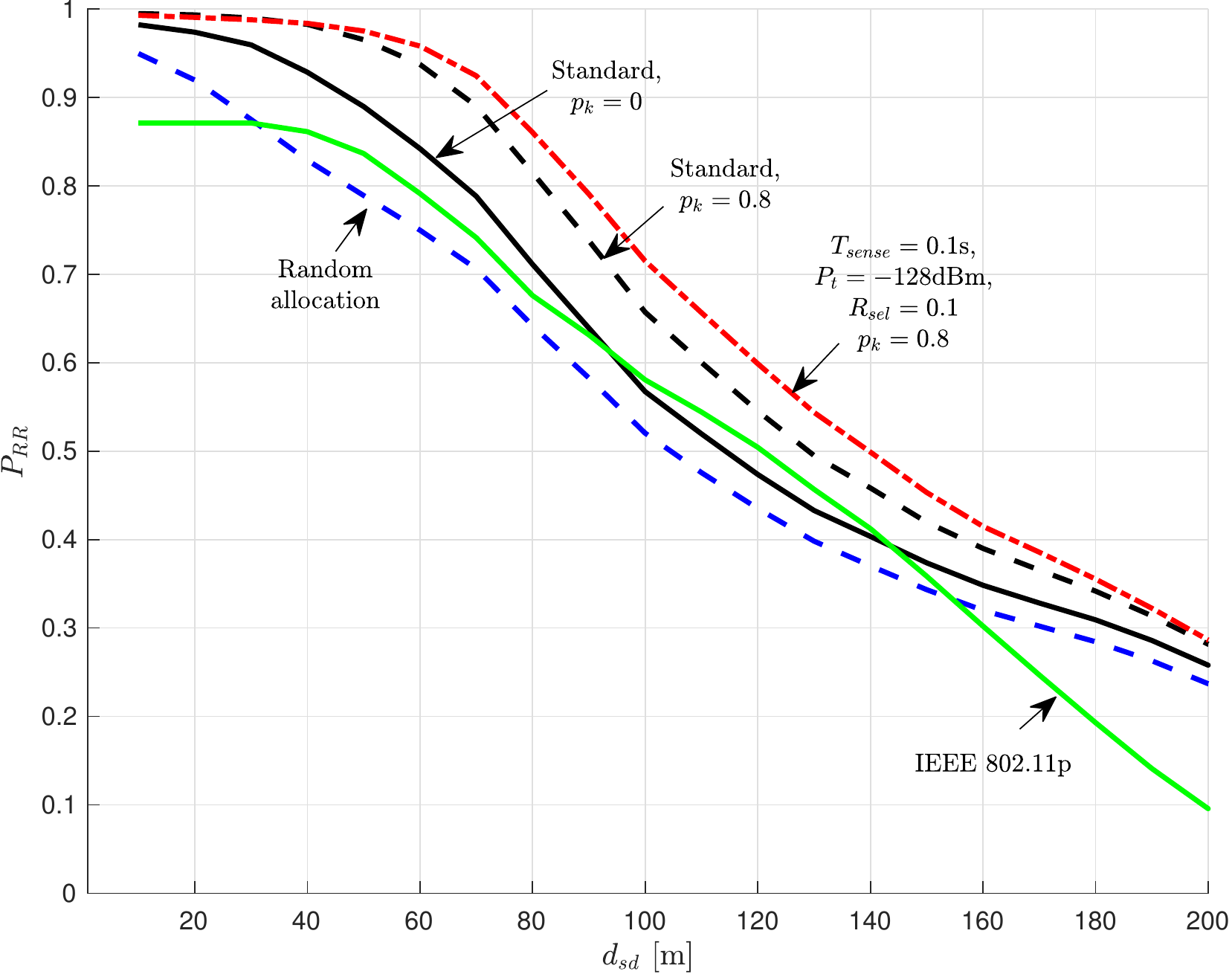}\label{fig:summaryPrrBologna}}
	\subfigure[Highway.]
	{\includegraphics[width=0.88\linewidth,draft=false]{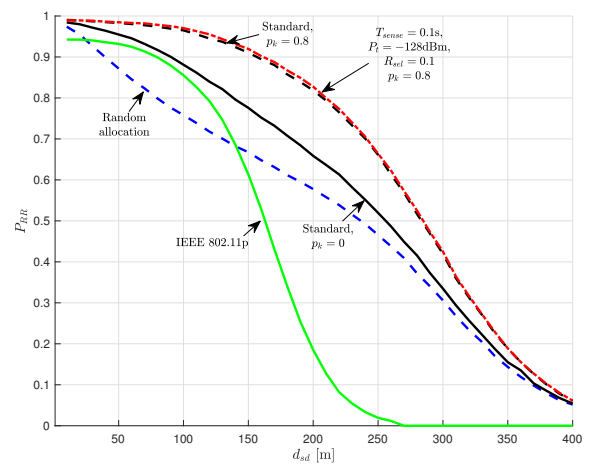}\label{fig:summaryPrrHighway}}
	\caption{Summary comparison in terms of \ac{PRR}. The values listed in Table~\ref{Tab:parameters} are used for those settings not explicitly indicated.}\label{Fig:summaryPrr}
\end{figure}

\begin{figure}[t!]
	\centering
	\subfigure[Cologne.]
	{\includegraphics[width=0.88\linewidth,draft=false]{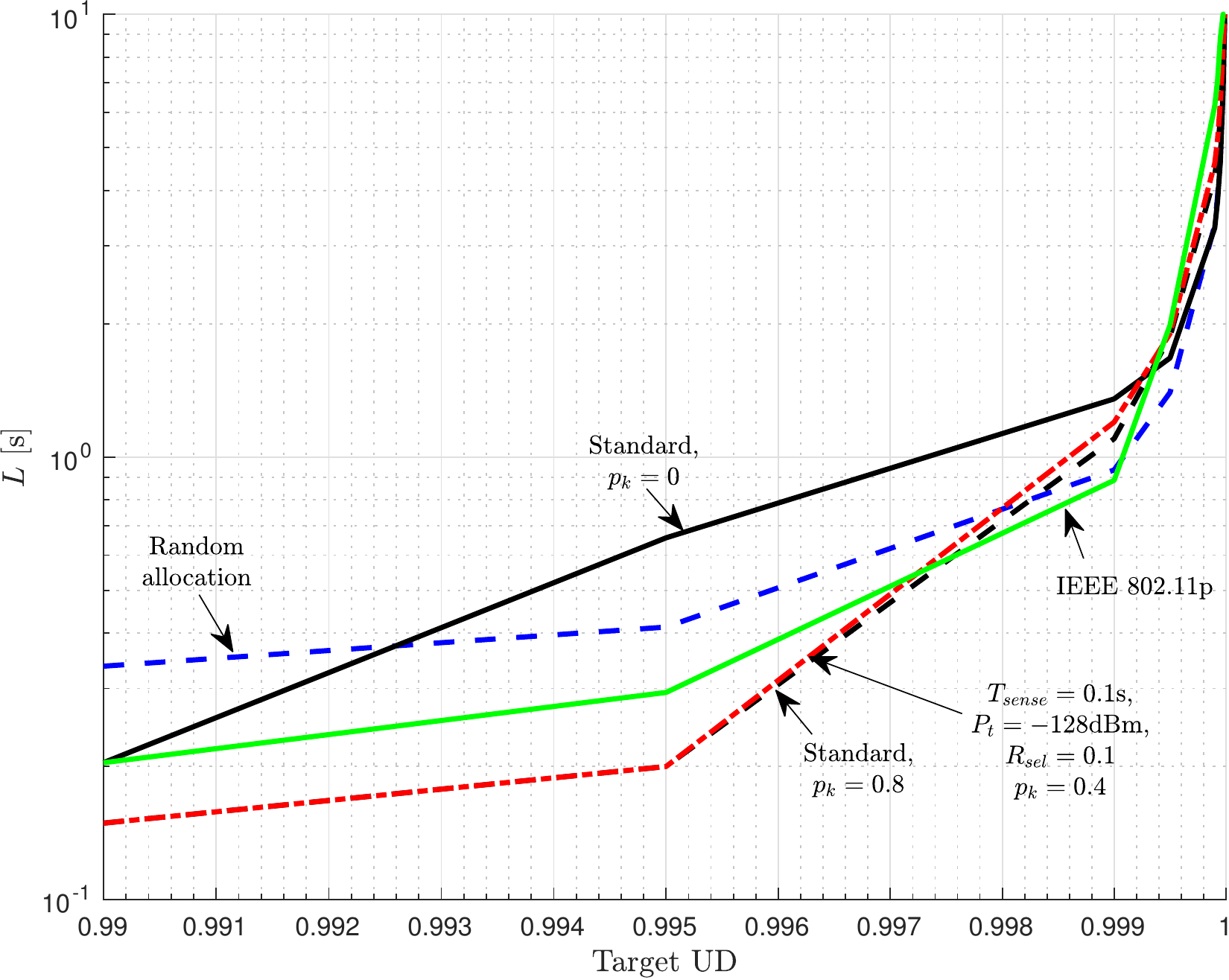}\label{fig:summaryUDColonia}}
	\subfigure[Bologna.]
	{\includegraphics[width=0.88\linewidth,draft=false]{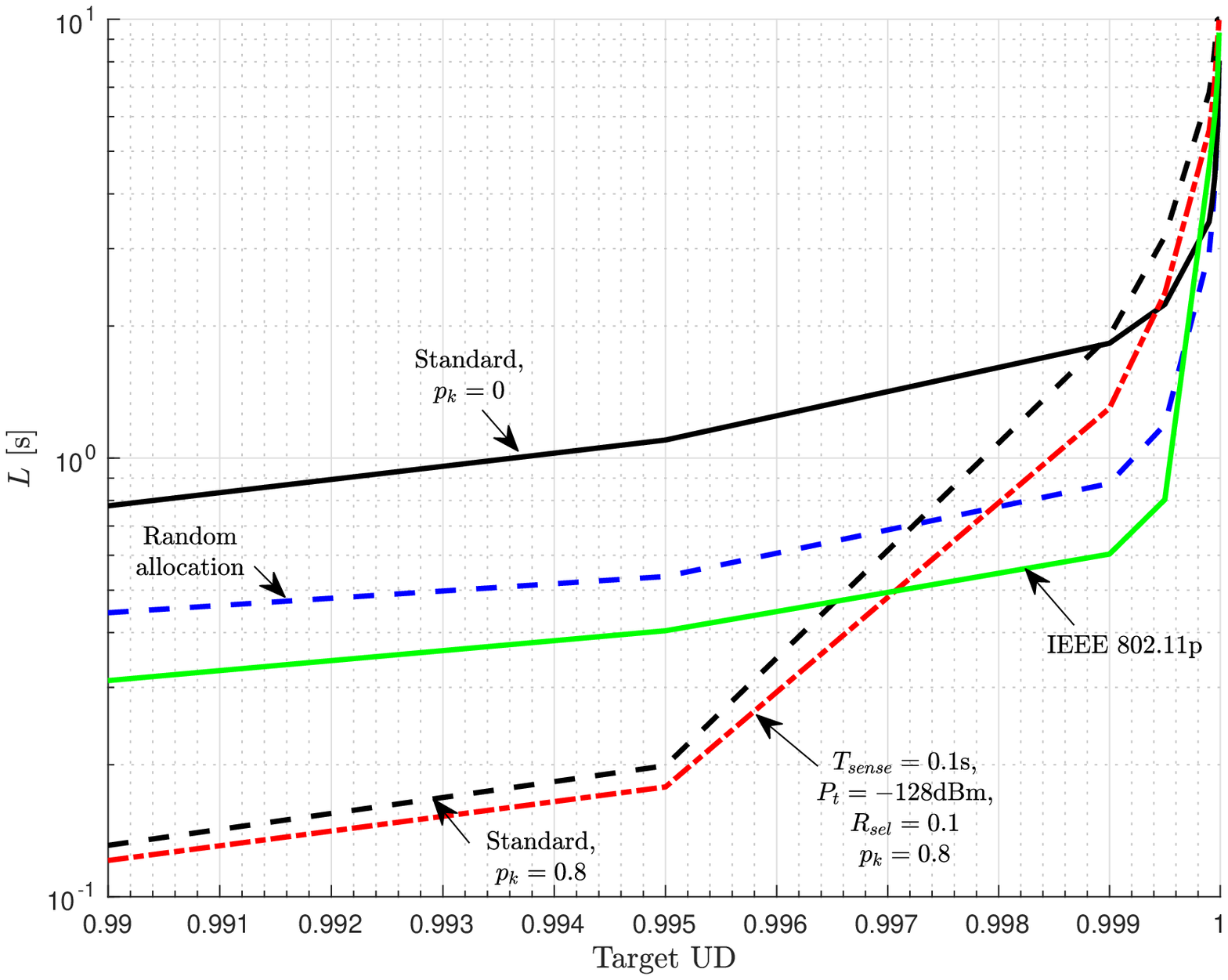}\label{fig:summaryUDBologna}}
	\subfigure[Highway.]
	{\includegraphics[width=0.88\linewidth,draft=false]{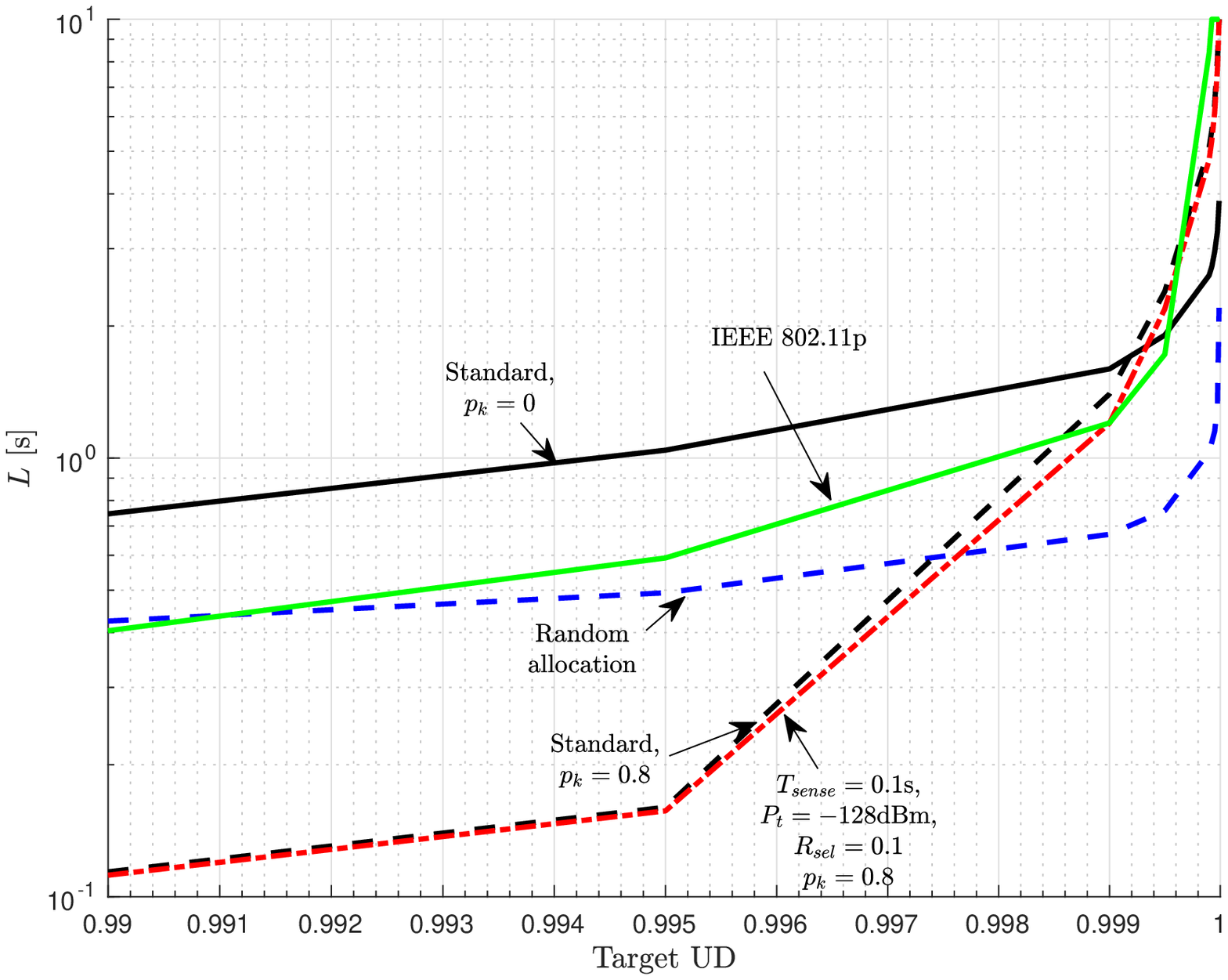}\label{fig:summaryUDHighway}}
	\caption{Summary comparison in terms of \ac{UD}.}\label{Fig:summaryUD}
\end{figure}

Focusing on the PRR in Fig.~\ref{Fig:summaryPrr}, thanks to the more advanced PHY and MAC protocols  LTE-V2V Mode~4 is shown to outperform IEEE 802.11p in almost all situations,  especially if $\pkeep=0.8$ is assumed in LTE. It is however to note that the gap between the technologies is not as large as presumable. The superiority of LTE-V2V is more debatable if we now focus on the UD in Fig.~\ref{Fig:summaryUD}. In some cases, such as with target 0.999 in Bologna, the update delay of IEEE 802.11p is indeed lower than with LTE; this is due to a significantly lower correlation between errors that follow the CSMA/CA protocol. 

\subsubsection{Random allocation} 
Using the random allocation as a reference, all curves in Fig.~\ref{Fig:summaryPrr} give evidence that Mode~4 is effective in the identification of the free resources and the consequent spacial reuse. At the same time, Fig.~\ref{Fig:summaryUD} illustrates how this comes at the cost of a higher correlation in the errors. If we focus as an example on the Highway scenario, when a target 0.999 or more is considered, Mode~4 has an update delay which is double or more than with the random allocation. 

\subsubsection{Standard Mode~4} Restricting the observation to Mode~4, it can be noted that $\pkeep=0.8$ allows to improve PRR at the cost of a higher UD. The improvement in terms of PRR (Fig.~\ref{Fig:summaryPrr}) is negligible in Cologne, where the scenario is sparse, but becomes clear in both Bologna and Highway. The impact in terms of UD (Fig.~\ref{Fig:summaryUD}) requires a more careful discussion: it can be noted, in fact, that $\pkeep=0.8$ allows also a lower UD until for values of the target percentile below a given threshold (between 0.997 and 0.998), but causes a higher UD above it. As discussed in Section~\ref{sec:MAC}, this behaviour is a consequence of longer intervals with the same allocation when $\pkeep$ is higher; with $\pkeep=0.8$, wrong estimation of occupied resources is less probable, with lower UD at low target percentile, but an error causes longer bursts of errors, implying higher UD at higher target percentiles.

\subsubsection{Mode~4 with Optimized Parameters} In Fig.~\ref{Fig:summaryPrr} and Fig.~\ref{Fig:summaryUD}, we also show the performance obtained by optimizing the Mode~4 parameters as discussed in the previous sections and summarized in Table~\ref{Tab:parameters}. The clear result is that some margin for optimization in terms of PRR is possible, even if small and limited to very crowded situations (i.e., in Bologna). The impact to the UD is not remarkable, with a slight improvement compared to the standard Mode~4 with $\pkeep=0.8$ in the congested scenarios and a negligible worsening in the Cologne scenario.

\subsubsection{Comparison with related work} The discussed results appear consistent with those presented in the related work. Specifically, focusing on a highway scenario, in \cite{TogSaiMugMahetal:AX18} the authors show that \ac{PRR} improves increasing $\pkeep$, which is in total agreement with Fig.~\ref{fig:summaryPrrHighway}. In \cite{NguShaSudKapEtc:C17}, LTE-V2V is shown to outperform IEEE~802.11p both in urban and highway scenarios; the gap is doubled in the latter case; these results are fundamentally coherent with Figs.~\ref{fig:summaryPrrColonia}, Figs.~\ref{fig:summaryPrrBologna}, and Figs.~\ref{fig:summaryPrrHighway}, especially if a high $\pkeep$ is assumed. Finally, in \cite{MolGoz:C17}, the authors show that with $\pkeep=0$ in a not congested urban
	scenario, Mode~4 provides a delivery rate not too much higher than a random allocation. Although we observed some improvement also in the (least loaded) Cologne scenario, the cited results appear compatible with the fact that reducing the density of vehicles from Highway to Bologna to Cologne, the gap between Mode~4 and a random allocation reduces (Fig.~\ref{Fig:summaryPrr}).	

\input{TableParameters_indications_clean.tex}

\section{Conclusion}\label{sec:conclusion}

In this work we have described the various parameters affecting the performance of the 3GPP Mode 4 algorithm of LTE-V2X. Results have been shown separating those acting at the PHY and MAC layer and performing simulations in three different realistic scenarios. The main conclusions, which also lead to the specific indications reported as the last column of Table~\ref{Tab:parameters_indications}, can be summarized as follows.
\begin{itemize}
	\item The modification of all parameters appears almost irrelevant in scenarios with a low to medium  number of vehicles, whereas it might become significant when congestions occur;
	\item Most parameters at the PHY layer have a minor impact on the performance and should be set to the extremes allowed by the specifications: specifically, the threshold power to sense the resource as occupied and the first \ac{TTI} for the next allocation should be set to the minimum, whereas the last \ac{TTI} for the next allocation should be set to the maximum;
	\item A minimum improvement is possible by reducing the portion of beacon resources passed to the MAC layer, presently fixed to 20\% by the specifications; it has also been observed that the best number slightly varies with the density of vehicles;
	\item Again at the PHY layer, some performance improvement can be achieved acting on the sensing period, which is currently fixed to 1~s; it however requires to rethink the way the channel is sensed, since just reducing it is against the fact that some nodes may be transmitting at 1~Hz;
	\item At the MAC layer, by modifying the keep probability it is possible to trade-off between a higher packet reception probability and a lower update delay; the variation can be relevant;
	\item The variability of the keep probability appears sufficient to control the system performance; thus, as defined in the specifications, it appears acceptable to have fixed values for the minimum and maximum number of beacon periods before a reallocation.
\end{itemize}

\section*{Appendix A: SINR Calculation}

Given a generic node $a$, $\mathcal{S}_a$ and $\mathcal{B}_a$ will be respectively used  to denote the set of subframes (we define a set for generality, although it will be always of one element in our case) and the portion of bandwidth where $a$ transmits.  
Given the generic transmitter $i$ and the generic receiver $j$, if $\mathcal{S}_i \cap \mathcal{S}_j \neq \emptyset$, $i$ cannot decode the message from $j$ due to half duplex limitations. Otherwise, the message is correctly decoded if the average \ac{SINR}, denoted as $\gamma_{ij}$, is higher than a given threshold $\gammamin$. The average \ac{SINR} is calculated as
\begin{equation}
\gamma_{ij} = \frac{\psi_{ij}}{P_\text{n} + \sum\limits_{\substack{k\in\mathcal{V}-\left\{i,j\right\}}}{K_{\text{S}}}(\mathcal{S}_k.\mathcal{S}_i) \; {K_{\text{IBE}}}(\mathcal{B}_k,\mathcal{B}_i) \;\psi_{kj}}
\end{equation}
where $\psi_{ab}$ is the power received by $b$ from $a$, $P_\text{n}$ is the average noise power, $\mathcal{V}$ is the set of all the vehicles in the scenario, ${K_{\text{S}}}(\mathcal{S}_a,\mathcal{S}_b)$ is the portion of time when the two signals from $a$ and $b$ overlap, and $K_{\text{IBE}}(\mathcal{B}_a,\mathcal{B}_b)$ is the \ac{IBE} coefficient from a signal transmitted in the frequency portion $\mathcal{B}_a$ to the frequency portion $\mathcal{B}_b$. ${K_{\text{S}}}(\mathcal{S}_a,\mathcal{S}_b)$ is proportional to the number of subframes they overlap, with 0 if the two signals use different subframes and 1 if they use the same subframes. The \ac{IBE} coefficient is calculated as detailed in \cite{3GPP_TS_36_101}, with a value of 1 if the signals overlap in frequency and a lower value otherwise.

\section*{Appendix B: Calculation of the Hidden Node Probability}

The hidden node probability shown in Fig.~\ref{Fig:hiddenNodes} is calculated as follows. We denote the set of all nodes at a given time instant as $\mathcal{N}$. Then, any node in the scenario $a\in\mathcal{N}$ is a potential source and we define as a generic destination $b$ any node that receives from $a$ with sufficient \ac{SNR}, thus any $b\in\mathcal{D}_a$, with $$\mathcal{D}_a = \left\{b\in\mathcal{N}-\{a\}\biggr|\frac{\psi_{a,b}}{P_\text{n}}>\gammamin\right\}$$ where $\psi_{x,y}$ is the power received by $y$ when $x$ is transmitting, $P_\text{n}$ is the noise power, and $\gammamin$ is a suitable threshold. Furthermore, we denote as $c$ the generic interfering node, which is any node that causes the \ac{SINR} to become lower than the given threshold $\gammamin$ (this excludes the nodes that are too far), which can be written as $c\in\mathcal{I}_{a,b}$, with $$\mathcal{I}_{a,b}=\left\{c\in\mathcal{N}-\{a,b\}\biggr|\frac{\psi_{a,b}}{P_\text{n}+\psi_{c,b}}<\gammamin\right\}\;.$$
	Among all interfering nodes, we call hidden any node $h$ that the source cannot hear, using the same threshold $\gammamin$ as a discriminator. In formulas, $h \in \mathcal{H}_{a,b}$, where $$\mathcal{H}_{a,b}=\left\{h\in\mathcal{I}_{a,b}\biggr|\frac{\psi_{h,a}}{P_\text{n}}<\gammamin\right\}\;.$$
	At a given instant, the hidden node probability $P_\text{hn}$ is calculated as 
	$$P_\text{hn} = \sum_{a\in\mathcal{N}}\sum_{b\in\mathcal{D}_a} \frac{\#\mathcal{H}_{a,b}}{\#\mathcal{I}_{a,b}}$$ where $\#\mathcal{X}$ is the cardinality of set $\mathcal{X}$.
	Finally, the overall hidden node probability is given by the average of $P_\text{hn}$ over all considered instants, which in our case correspond to periodic samples of period equal to the beacon interval, i.e., $\Tbeacon=100$~ms.

\section*{Appendix C: Demonstration of Equation (3)}

Equation \eqref{eq:PTkeep} gives the probability that the same allocation is kept for a given number of beacon periods, i.e., the probability that a \ac{TBC} lasts for a given number of $\Tbeacon$.

Let us denote as $\NTBE$ the number of \ac{TBE} of which the generic \ac{TBC} is composed. The probability distribution $\Pntotcond(n,1)$ that the same allocation is kept for a given number $n$ of beacon periods, conditioned to the fact that the \ac{TBC} is composed by a single \ac{TBE}, is simply
\begin{align}\label{eq:Pntotcond1}
\Pntotcond(n,1)=\Prob\{\Pntot(n)|\NTBE=1\} = \Pnsingle(n)\;.
\end{align}

Since the length of each \ac{TBE} is independent to the others and given the definition of $\pvector$, the probability distribution $\Pntot(n|2)$ that the same allocation is kept for a given number $n$ of beacon periods, conditioned to the fact that the \ac{TBC} is composed by two \ac{TBE}, becomes
\begin{align}\label{eq:Pntotcond2}
\Pntotcond(n,2)&=\Prob\{\Pntot(n)|\NTBE=2\} = \Pnsingle(n) * \Pnsingle(n) \nonumber\\ &= \InvFouD{ \FouD{\pvector} \cdot \FouD{\pvector}}
\end{align}
where the symbol $*$ denotes the convolution operation and the symbol $\cdot$ means the multiplication element-by-element of the vectors. The use of discrete Fourier transformations allows to convert convolutions into products. Straightforwardly, \eqref{eq:Pntotcond2} can be generalized, for a generic number of \ac{TBE} composing the \ac{TBC} as 
\begin{align}\label{eq:PntotcondN}
\Pntotcond(n,k)&=\Prob\{\Pntot(n)|\NTBE=k\} \nonumber\\ &= \InvFouD{ \left(\FouD{\pvector}\right)^k}
\end{align}
where the exponent to the vector denotes its application to each element.

Then, the probability that the \ac{TBC} includes $k$ \ac{TBE}, denoted as $\PkTBE(k)$, is equal to the probability to keep it $k-1$ times and do not keep it the last one, i.e., 
\begin{align}\label{eq:PkTBE}
\PkTBE(k) = \Prob\{\NTBE=k\} = \pkeep^{k-1} (1-\pkeep)\;.
\end{align}

Using \eqref{eq:PntotcondN} and \eqref{eq:PkTBE} and summing up for a variable number of \ac{TBE}, it follows \eqref{eq:PTkeep}.

\section*{Appendix D: Demonstration of Equation (4)}

Equation \eqref{eq:probChange} corresponds to the probability that a change occurs within a given number of beacon intervals, i.e., the probability that a \ac{TBC} ends within an interval of the given number of $\Tbeacon$.

Let us focus on the last \ac{TBC} that starts before the given observation interval. The probability to be calculated is indeed exactly equal to the probability that such \ac{TBC} ends within the observation interval. Let us use $l$ to denote the length (in number of beacon intervals) of the \ac{TBC} and $w$ that of the observed interval. The probability that a change occurs within $w$ is obviously 1 if $l\leq w$. If $l > w$, it depends on when it started: the change occurs if the interval does not start in the last $l-w$ beacon intervals, which means, assuming equal probability that it started on a generic beacon interval before the observation interval not farther than $w$ beacon periods, that a change occurs with probability $(l-w)/l$. 

Summarizing, the probability $\pCambcond(w,l)$  to observe a change within the observation interval of length $w$, conditioned to having the length of \ac{TBC}, denoted as $\TTBC$ and expressed in beacon intervals, equal to $l$ is
\begin{equation}\label{eq:pCambcond}
\pCambcond(w,l)= \Prob\{\pCamb(w)|\TTBC=k\} = \left \{
\begin{array}{ll}
1 & w \leq=l \\
\frac{l-w}{l} & \text{otherwise} \\
\end{array} 
\right.\;.
\end{equation}

The probability that a change occurs within the observation interval can then be easily calculated, using \eqref{eq:pCambcond} and  \eqref{eq:PTkeep}, as the probability that no change occurs later than the given observation interval, leading to \eqref{eq:probChange}.

\bibliographystyle{IEEEtran}
\bibliography{biblioSelf,biblioOthers,biblioStandards}

\begin{IEEEbiography}[{\includegraphics[width=1in,height=1.25in,clip,keepaspectratio]{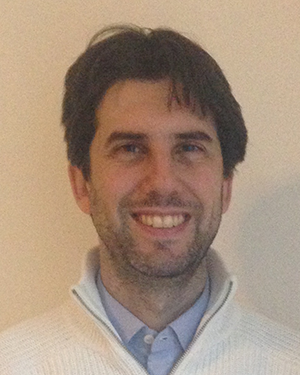}}]{Alessandro Bazzi} (S'03-M'06-SM'18) received the Laurea degree and the Ph.D. degree in telecommunications engineering both from the University of Bologna, Italy, in 2002 and 2006, respectively. Since 2002, he works with the Institute of Electronics, Computer and Telecommunication Engineering (IEIIT) of the National Research Council of Italy (CNR) and since the academic year 2006/2007, he has been acting as adjunct Professor at the University of Bologna. His work mainly focuses on connected vehicles and heterogeneous wireless access networks, with particular emphasis on medium access control, routing and radio resource management. Dr. Bazzi serves as a Reviewer and TPC Member for various international journals and conferences and he is currently in the Editorial Board of Hindawi Mobile Information Systems and Hindawi Wireless Communications and Mobile Computing.
\end{IEEEbiography}

\begin{IEEEbiography}[{\includegraphics[width=1in,height=1.25in,clip,keepaspectratio]{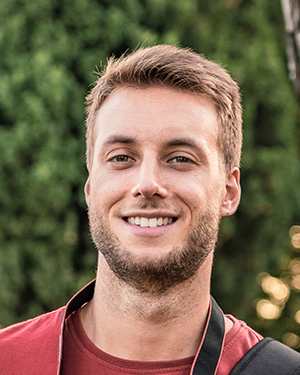}}]{Giammarco Cecchini} (S'17) received the Laurea degree (summa cum laude) in Telecommunications Engineering in Bologna in 2016, awarded with a prize from Lions Club Fondazione Guglielmo Marconi for his master thesis entitled "Performance of LTE-V2V for Cooperative Awareness". Since 2017, he joined the Institute of Electronics, Computer and Telecommunication Engineering (IEIIT) of the National Research Council of Italy (CNR). His work mainly focuses on wireless technologies for connected vehicles, both including IEEE 802.11p (with related standards) and LTE-V2X. He is also the main developer of the open-source MATLAB simulator LTEV2Vsim.
\end{IEEEbiography}

\begin{IEEEbiography}[{\includegraphics[width=1in,height=1.25in,clip,keepaspectratio]{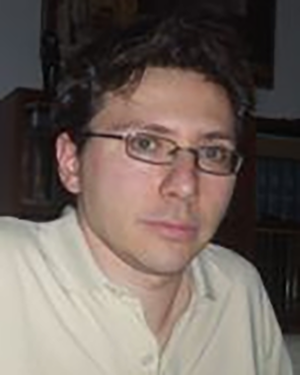}}]{Alberto Zanella} (S'99-M'00-SM'12) received the Laurea degree (summa cum laude) in Elecronic Engineering from the University of Ferrara, Italy, in 1996, and the Ph.D. degree in Electronic Engineering and Computer Science from the University of Bologna in 2000. In 2001 he joined the CNR-CSITE (merged in CNR-IEIIT since 2002) as a researcher and, since 2006, as senior researcher. His research interests include MIMO, mobile radio systems, ad hoc and sensor networks, vehicular networks. Since 2001 he has the appointment of Adjunct Professor of Electrical Communication (2001 - 2005), Telecommunication Systems (2002-2012-2013), Multimedia Communication Systems (2006 - 2011) at the University of Bologna. He participated/participate to several national and European projects. He was Technical Co-Chair of the PHY track of the IEEE conference WCNC 2009 and of the Wireless Communications Symposium (WCS) of IEEE Globecom 2009. He was/is in the Technical Program Committee of several international conferences, such as ICC, Globecom, WCNC, PIMRC, VTC. He had served as Editor for Wireless Systems (2003-2012), IEEE TRANSACTIONS ON COMMUNICATIONS. and he is currently Senior Editor for the same journal.
\end{IEEEbiography}

\begin{IEEEbiography}[{\includegraphics[width=1in,height=1.25in,clip,keepaspectratio]{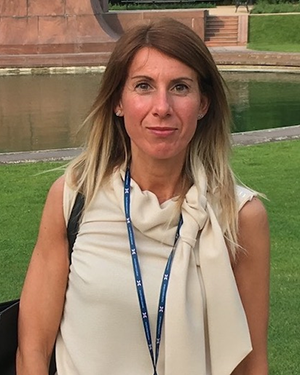}}]{Barbara M. Masini} (S'02-M'05) received the Laurea degree (summa cum laude) in Telecommunications Engineering and the Ph.D. degree in Electronic, Computer Science, and Telecommunication engineering from the University of Bologna, Italy, in 2001 and 2005, respectively. Since 2005, she is a researcher at the Institute for Electronics and for Information and Telecommunications Engineering (IEIIT), of the National Research Council (CNR). Since 2006 she is also adjunct Professor at the University of Bologna. She works in the area of wireless communication systems and her research interests are mainly focused on 5G vehicular networks, facing new challenges from physical and MAC aspects up to the applications and real field trial implementations. Research is also focused on relay assisted communications, energy harvesting and on theoretical and experimental activities dealing with visible light communication. She is Editor of Elsevier Computer Communication. She is Secretary of Chapter VT06/COM19 of the IEEE Italy section. She is TPC member of several conferences, reviewer for most international journals and for the Italian Ministry of Economic Development (MISE).
\end{IEEEbiography}
\EOD

\end{document}

%% file: definitions.tex
\def\Prob {{\mathbb P}}

\newcommand{\Tsense}{T_\text{sense}}
\newcommand{\Pthr}{P_\text{th}}
\newcommand{\PercR}{R_\text{sel}}
\newcommand{\nCandidates}{n_R}
\newcommand{\Tone}{T_\text{1}}
\newcommand{\Ttwo}{T_\text{2}}

\newcommand{\nMin}{n_\text{min}}
\newcommand{\nMax}{n_\text{max}}
\newcommand{\pkeep}{p_\text{k}}

\newcommand{\bBytes}{B}
\newcommand{\fBeacon}{f_\text{B}}
\newcommand{\Tbeacon}{T_\text{B}}
\newcommand{\nRes}{R}
\newcommand{\gammamin}{\gamma_\text{min}}

\newcommand{\FouD}[1]{\mathcal{F}_\text{d}\left\{{#1}\right\}}
\newcommand{\InvFouD}[1]{{\mathcal{F}_\text{d}^{-1}}\left\{{#1}\right\}}

\newcommand{\pCamb}{P_\text{r}}
\newcommand{\pCambcond}{P_\text{r}^\text{cond}}
\newcommand{\Pntot}{P_\text{TBC}}
\newcommand{\Pntotcond}{{P_\text{TBC}^\text{cond}}}
\newcommand{\Pnsingle}{{P_\text{TBE}}}
\newcommand{\dist}{d_\text{sd}}
\newcommand{\pvector}{\mathbf{P}_\text{TBE}}
\newcommand{\PkTBE}{P_\text{kTBE}}
\newcommand{\TTBC}{D_\text{TBC}}
\newcommand{\NTBE}{N_\text{TBE}}

%% file: acronyms.tex
\begin{acronym} 
\acro{5G}{fifth-generation} 
\acro{AWGN}{additive white Gaussian noise}
\acro{BEP}{beacon error probability}
\acro{BP}{beacon periodicity}
\acro{BR}{beacon resource}
\acro{BSM}{basic safety message}
\acro{BSS}{basic service set}
\acro{C-ITS}{cooperative intelligent transport systems}
\acro{C-V2X}{cellular-\acl{V2X}}
\acro{CAM}{cooperative awareness message}
\acro{CAM-R}{\ac{CAM} resource}
\acro{CAV}{cooperative and automated vehicle}
\acro{CCA}{clear channel assessment}
\acro{ccdf}{complementary cumulative distribution function}
\acro{CD}{collision detection}
\acro{CDF}{cumulative distribution function}
\acro{c.f.}{characteristic function}
\acro{CSI}{channel state information}
\acro{CSMA/CA}{carrier sense multiple access with collision avoidance}
\acro{D2D}{device-to-device}
\acro{DSRC}{dedicated short range communication}  
\acro{ERP}{equivalent radiated power}
\acro{FCD}{floating car data}
\acro{FD}{full duplex}
\acro{FDD}{frequency division duplex}
\acro{GPS}{global positioning system}
\acro{HD}{half duplex}
\acro{i.i.d.}{independent identically distributed}
\acro{IBE}{in-band emission}
\acro{LOS}{line of sight}
\acro{LTE}{long term evolution}  
\acro{LTE-D2D}{\ac{LTE} with device-to-device communications}  
\acro{LTE-V2V}{\ac{LTE}-vehicle-to-vehicle}  
\acro{LTE-V2X}{\ac{LTE}-vehicle-to-anything}  
\acro{MAC}{medium access control}
\acro{MCS}{modulation and coding scheme}
\acro{MD}{allocation with maximum reuse distance}
\acro{NHTSA}{National Highway Traffic Safety Administration}
\acro{NLOS}{non-\ac{LOS}}
\acro{OBU}{on board unit}
\acro{OCB}{outside the context of a BSS}
\acro{OFDM}{orthogonal frequency division multiplexing}
\acro{PDF}{probability density function}
\acro{PEP}{pairwise error probability} 
\acro{PHY}{physical}
\acro{PMF}{probability mass function}
\acro{PPP}{Poisson point process}
\acro{ProSe}{proximity services}
\acro{PRR}{Packet reception ratio}
\acro{PSCCH}{physical sidelink control channel}
\acro{QoS}{quality of service}
\acro{r.v.}{random variable}
\acro{RF}{radio frequency} 
\acro{RR}{random resource allocation}
\acro{RSRP}{reference signal received power}
\acro{RSU}{road side unit} 
\acro{SC-FDMA}{single carrier-frequency division multiple access}
\acro{SHINE}{simulation platform for heterogeneous interworking networks}
\acro{SI}{self-interference}
\acro{SINR}{signal to noise and interference ratio}
\acro{SNR}{signal to noise ratio}
\acro{SPS}{semi-persistent scheduling}
\acro{S-RSSI}{sidelink-received signal strength indicator}
\acro{SC-FDMA}{single carrier-frequency division multiple access}
\acro{SCI}{sidelink control information}
\acro{SV}{smart vehicle}
\acro{TB}{transport block}
\acro{TBE}{time before evaluation}
\acro{TBC}{time before change}
\acro{TDD}{time division duplex}
\acro{TTI}{transmission time interval}
\acro{UD}{Update delay}
\acro{V2C}{vehicle-to-cellular} 
\acro{V2I}{vehicle-to-infrastructure} 
\acro{V2P}{vehicle-to-pedestrian}
\acro{V2R}{vehicle-to-roadside}
\acro{V2V}{vehicle-to-vehicle} 
\acro{V2X}{vehicle-to-anything} 
\acro{VANETs}{vehicular ad hoc networks} 
\acro{WAVE}{wireless access in vehicular environment}
\end{acronym}

%% file: TableParameters_values_clean.tex
\begin{table*}[t]
\caption{Main parameters of Mode 4, with constraints indicated by 3GPP and values used where not differently specified.}
\vspace{-2mm}
\label{Tab:parameters}
\centering
\footnotesize
\begin{tabular}{p{10cm}p{3cm}p{2.4cm}}
 & \textbf{Constraints from 3GPP} & \textbf{Used if not specified} 
 \\ \hline
\textbf{PHY} & \\
Sensing period ($\Tsense$)& 1\;s & 1\;s (mandated)\\
Minimum threshold to the power level ($\Pthr$) & $\in[-128,-2]$ dBm & -110 dBm (Ref.~\cite{MolGoz:J17}) \\
Portion of beacon resources passed to the MAC ($\PercR$) & 0.2 & 0.2 (mandated) \\
First subframe for the next allocation ($\Tone$) & $\leq$4 & 1 (lowest) \\
Last subframe for the next allocation ($\Ttwo$) & $\geq$20, $\leq$100 & 100 (highest) \\
\vskip 0cm
\textbf{MAC} & & \\
Minimum number of beacon periods before evaluating a new reallocation ($\nMin$) & 5 & 5 (mandated) \\
Maximum number of beacon periods before evaluating a new reallocation ($\nMax$) & 15 & 15 (mandated) \\
Probability to keep the same resource ($\pkeep$) & $\in [0,0.8]$ & 0.4 (intermediate)\\
\hline
\end{tabular}
\end{table*}

%% file: TableSettings.tex
\begin{table}[t]
\caption{Main settings.}
\vspace{-2mm}
\label{Tab:Notations}
\centering
\footnotesize
\begin{tabular}{p{3.5cm}p{1.1cm}p{1.3cm}p{1.3cm}}
\hline
\textbf{Scenario and application} & & &  \\
Scenario & Cologne & Bologna & Highway \\
Size & 3.4 km$^2$ & 2.1 km$^2$ & 16 km, \\
& & & 3+3 lanes \\
Average n. of vehicles & 925 & 667 & 2015 \\
Reference awareness distance & 100 m & 100 m & 200 m \\
Average neighbours (std. dev.) & 14.8 (8.8) & 25.4 (25.4) & 49.4 (12.5) \\
Beacon periodicity  & 10 Hz & & \\
Beacon size & 300~bytes & & \\  
\vskip 0cm
\textbf{PHY settings} &  & & \\
Bandwidth & 10 MHz & & \\
Transmission power & 23~dBm & & \\
Antenna gain (both tx and rx)  & 3 dB & & \\
Noise figure & 9~dB & & \\
Propagation model & \multicolumn{3}{l}{WINNER+, Scenario B1} \\
Shadowing variance & \multicolumn{3}{l}{LOS 3 dB, NLOS 4 dB}\\
MCS & \multicolumn{3}{l}{4 in Cologne (1 BR/TTI)} \\
& \multicolumn{3}{l}{7 in Bologna \& Highway (2 BRs/TTI)} \\
Minimum SINR & \multicolumn{3}{l}{MCS 4: 2.76~dB, MCS 7: 7.30~dB} \\
\hline
\end{tabular}
\end{table}

%% file: TableParameters_indications_clean.tex
\begin{table*}
\caption{Main conclusions per each parameter.}
\vspace{-2mm}
\label{Tab:parameters_indications}
\centering
\footnotesize
\begin{tabular}{p{5cm}p{2.9cm}p{9cm}}
 & \textbf{Constraints by 3GPP} & \textbf{Summary indications} \\
\hline
\textbf{PHY} & & \\
Sensing period & $\Tsense=1$\;s & The lower it is and the more accurate is the estimation, but it is constrained by the minimum beacon period  \\
Minimum threshold to the power level & $\Pthr\in[-128,-2]$ dBm & The lower is better, since the used resources are better individuated \\
Portion of resources passed to the MAC & $\PercR=0.2$ & 0.1 might be slightly better in very congested scenarios: it limits the MAC choice to the less interfered resources, without reducing the set too much \\
First/Last subframes for the next allocation & $\Tone\leq4$, $\Ttwo\in\{20,100\}$ & Negligible impact in all investigated scenarios. Expected anyway that the longer is the interval and the better it is \\
\vskip 0cm
\textbf{MAC} & & \\
Min/Max beacon periods before evaluating a reallocation & $\nMin=5$, $\nMax=15$ & Modifications to $\pkeep$ have a similar impact \\
Probability to keep the same resource & $\pkeep\in [0,0.8]$ & PRR/UD trade-off: if $ \pkeep \uparrow$ then PRR $\uparrow$ and UD $\uparrow$ \\
\hline
\end{tabular}
\end{table*}